%% file: hiders.tex
\documentclass[preprint]{elsarticle}

\input{macros}

\journal{Information Sciences}

\begin{document}


\title{Hiders' Game}

\author[1]{Gleb Polevoy\corref{cor1}%
\fnref{fn1}}
\ead{gpolevoy@mail.uni-paderborn.de}

\author[2]{Tomasz Michalak
}
\ead{tpm@mimuw.edu.pl}

\cortext[cor1]{The corresponding author.}
\fntext[fn1]{Much of this work was performed when the author was at the University of Warsaw.}

\address[1]{University of Paderborn,
Fürstenallee 11,
33102,
Paderborn,
Germany}

\address[2]{University of Warsaw,
Banacha 2,
02-097,
Warsaw,
Poland}

\input{abstract}









\maketitle

\input{introduction}
\input{model}
\input{hiders_game}

\input{hiders_cases}
\input{related_work}
\input{advice}

\input{conclusion}




\bibliographystyle{elsarticle-num}
\bibliography{library}

\newpage
\appendix

\input{omitted_proofs}

\input{non_appl_proofs}

\end{document}

%% file: macros.tex
\usepackage[linesnumbered,ruled]{algorithm2e}

\usepackage{url}
\usepackage{todonotes}

\usepackage{latexsym,amsmath,amssymb}
\usepackage{amsthm}
\usepackage{graphicx}

\usepackage{ifthen}
\usepackage{mdwlist,paralist}

\usepackage{comment}
\excludecomment{longversion}

\usepackage{epsfig}
\usepackage{epsf}

\usepackage[geometry]{ifsym}
\usepackage{rotating}

\usepackage{float}

\usepackage[utf8]{inputenc}
\usepackage{tikz}
\usepackage[leftcaption]{sidecap}
\sidecaptionvpos{figure}{m}
\usetikzlibrary{arrows}

\usepackage{paralist}

\newboolean{Longversion} \setboolean{Longversion}{false}

\newcommand*{\eqns}[1]{Eq.~#1}
\newcommand*{\eqnsref}[1]{\eqns{\eqref{#1}}} 

\newcommand*{\sectn}[1]{Section~#1}
\newcommand*{\sectnref}[1]{\sectn{\ref{#1}}} 
\newcommand*{\fig}[1]{Figure~#1}
\newcommand*{\figref}[1]{\fig{\ref{#1}}} 

\newcounter{subsecinl}[section]
\newcommand{\subsectioninline}[1]{\par\refstepcounter{subsecinl}%
\vspace{0.1cm} \noindent \textbf{\arabic{section}.\arabic{subsecinl} #1:}}

\newcommand{\anote}[1]{}

\newcommand{\set}[1]{\left\{ #1 \right\}} 
 
\newcommand{\paren}[1]{\left( #1 \right)}

\newcommand{\abs}[1]{\left| #1 \right|}

\newcommand*{\defas}{\ensuremath{\stackrel{\rm \Delta}{=}}}
\newcommand{\opt}{\textsc{opt}}
\DeclareMathOperator*{\Neighb}{N}

\newtheorem{theorem}{Theorem}
\newtheorem{defin}{Definition}
\newtheorem{remark}{Remark}
\newtheorem{lemma}{Lemma}
\newtheorem{proposition}{Proposition}

\newtheorem{corollary}{Corollary}
\newtheorem{observation}{Observation}
\newtheorem{example}{Example}

\newcommand{\defined}[1]{{\bf\emph{#1}}}

\newcommand{\NE}{\text{NE}}
\newcommand{\NS}{NS} 
\newcommand{\SPE}{SPE} 

\newcommand{\PANE}{PANE} 
\newcommand{\PANS}{PANS} 

\DeclareMathOperator{\cnct}{\mathit{Con}} 
\DeclareMathOperator{\dcnct}{\mathit{DisCon}} 
\DeclareMathOperator{\poa}{PoA} 
\DeclareMathOperator{\pos}{PoS} 

\DeclareMathOperator{\sw}{SW} 



%% file: abstract.tex
\begin{abstract}
Consider spies infiltrating a network or dissidents secretly organising
under a dictatorship.
Such scenarios can be cast as adversarial social network analysis problems involving nodes connecting while evading network analysis tools, e.g., centrality measures or community detection algorithms. While most works consider unilateral actions of an evader, we define a network formation game. Here, several newcomers attempt to rewire the existing social network, to become directly tied with the high centrality players, while keeping their own centrality small. This extends the network formation literature, including the Jackson and Wolinsky model, by considering additional strategies and new utility functions.
We algorithmically demonstrate that the pairwise Nash stable networks (\PANS) constitute a lattice, where the stronger \PANS{} lattice is nested in the weaker \PANS. We also prove that inclusion in \PANS{} implies less utility for everyone. Furthermore, we bound the social efficiency of \PANS{} and directly connect efficiency to the strength of \PANS.
Finally, we characterise the \PANS{} in
practically important settings, deriving tight efficiency bounds. 
Our results suggest the hiders how to interconnect stably and efficiently. Additionally, the results 
let us detect infiltrated networks,
enhancing the 
social network analysis tools. Besides the theoretical development, this is applicable 
to fighting terrorism and espionage. 
\end{abstract}



%% file: introduction.tex
\section{Introduction}
Oftentimes, network agents prefer to hide their cooperation and involvement. For instance, terrorists strive to obfuscate their leadership structure from security forces~\cite{ressler2006social}, and political dissidents hide from an authoritarian government~\cite{chase2002you,han2018contesting}. 
Lobbying foreign or merely unpopular interests~\cite{Choate1991} needs secrecy too. Unions and economical cooperators may at first want to avoid
publicity. Also software agents can prefer to hide, either for secrecy or due to the desire to avoid direct connection with potentially harmful agents. Such topics inspired
a recent increase in the literature which studied variations of the adversarial (or strategic) social network analysis problem~\cite{MichalakRahwanWooldridge2017,aziz2017weakening}. 
This problem involves network members ---e.g.~leaders of a covert network, or simply privacy-concerned people who want to gain information ---attempting to evade social network analysis tools by rewiring the network. Journalists may link with well-connected people for interviews and insides, while minimising their own visibility to avoid interference in their work. 
Activists can amplify their voice through connecting to influencers with similar views, and hiding helps avoiding opponents. Aspiring entrepreneurs may increase their exposure by connecting with key industry figures, while 
keeping a low profile, so as not to seem like a competition.
In this spirit, Waniek et al.~\cite{WaniekMichalakRahwanWooldridge2018} studied how a well-connected node can rewire the network to decrease her centrality but without sacrificing too much of her influence. These and other authors also considered how some communities could evade community detection algorithms~\cite{chen2019ga,zhou2019adversarial}. In both cases, 
simple heuristics turned out to be surprisingly effective in practice. 
Furthermore, evading link prediction algorithms~\cite{zhou2019attacking,zhou2019adversarial} have been studied too.

While the above works focused on analysing the possible actions of the (centralised) evader(s), in this paper, we propose the first study of the adversarial social network analysis problem as a \textit{game of network formation} between multiple evaders. We model scenarios, in which a group of newcomers attempt to connect to and/or possibly rewire the social network with the aim to become directly tied with nodes of high degree centrality while keeping their own centrality small.
Our objective is to understand what network topologies emerge in stable states of such a game, and use this for practical insights. 

Consider companies $A$ and $B$, both trying to steal the secrets of $C$ through industrial intelligence~\cite{bottom1984industrial}. To this end, each company engages moles to infiltrate company $C$. Each infiltrator aims to directly connect to the key people in $C$, while avoiding attracting attention herself. Similarly, the police would like their informants to have access to the members of the higher echelon of a criminal organisation but without being in the spotlight themselves~\cite{lindemann2017collaboration}.
Considering the degree centrality instead of the willingness of each given agent to disclose particular information assumes uniform passing of information; modelling strategic information transfer might sometimes increase the realism of the model, but would render the model parameter-sensitive and decrease its applicability.

We model such scenarios following the footsteps of the network formation literature~\cite{Jackson2003}. 
That is, we construct a non-cooperative game 
in which the players try to connect with each other and to the rest of the network strategically. While the players are active, the rest of the network is static, i.e., it is composed of nodes that do not participate in the game; we will refer to those as non-players. Non-players can represent na\"ive participants or simply machines.
The crucial element of our model are payoffs that reflect the desire of the players to be directly tied with the power/influence centers of the network but not to occupy exposed positions by themselves. Specifically, any player in our game aims to maximise the utility function that depends positively on the degree of her neighbours but negatively on her own degree. 

We consider networks that emerge as stable states of the above \textit{hiders' game}. To this end, following the network formation literature~\cite{BlochJackson2006}, we employ the concept of \textit{pairwise Nash equilibrium}. It is a refinement of the Nash equilibrium concept for networks that demands that any couple of players who would both benefit from connecting to one another, shall indeed become connected, as would indeed occur in many scenarios. This allows to avoid many unreasonable equilibria that would  otherwise arise. Nevertheless, our model differs from the previous network formation literature and, in particular, from the seminal model by Jackson and Wolinsky~\cite{JacksonWolinsky1996}, in some key respects. Firstly, the objective of our players is different as they aim to minimise their centrality and not maximise it, all while maximising connectedness to
high-centrality nodes, which is a proxy for power or influence. Secondly, we extend the set of possible strategies of the players by allowing an evader to actively
acquaint any two of her non-player neighbours and thereby interconnect them. This corresponds to an option to introduce one's neighbours to each other or to make their paths cross that is available in real life and on major social networking sites such as Facebook.

We analyse the (usual and strong) pairwise Nash equilibria emerging in the hiders' game and study their efficiency in terms of the prices of anarchy and stability.
The main results are summarised in Table~\ref{tab:our_main_res}. In particular, in \sectnref{Sec:hiders_game}, we demonstrate that pairwise Nash Equilibria exist and the resulting networks constitute a lattice, ordered by edge inclusion. The provided algorithms allow for computing
the edge-inclusion largest/smallest pairwise equilibria
and largest included/smallest including pairwise
equilibrium of any two given pairwise equilibria.
We prove that if an equilibrium edge-includes another profile, then all the players attain a higher utility in the including equilibrium. Together with the algorithmic results, this allows finding pairwise equilibria with the largest and with the smallest utility, which we also fully characterise. 
This leads to concrete policy recommendations both for the hiders and the seekers.
Additionally, we show that a pairwise $k$-strong equilibrium is strong if and only if each player obtains there her maximum utility among all the pairwise $k$-strong equilibria. Furthermore, the utility of all the pairwise $k$-strong equilibria is the
same if and only if each pairwise $k$-strong equilibrium
is actually strong. These equivalences allow for strength checking
based on efficiency and vice versa. We show that no general multiplicative efficiency bound exists, while providing an additive bound. We thus resort to concrete scenarios to obtain tighter bounds. 

In \sectnref{Sec:hiders_cases}, we first fully characterise the pairwise
Nash equilibria and their efficiency for the case of equal aversion
to having many neighbours, and then we characterise the case of
all the players besides one possessing a small aversion of many neighbours. These characterisations allow predicting all the possible pairwise equilibria, knowing the identities
of the hiding players, or guess who are the 
hiding players in a network, when these are undercover.
%
\sectnref{Sec:polic_rec} offers policy recommendations,
and we discuss limitations of our model and future directions
in \sectnref{Sec:conclusion}.
Throughout the paper, we defer some proofs to~\ref{Sec:omit_proof},
and we discuss the reason for proving that the equilibria constitute a lattice constructively in \ref{Sec:non_appl_proofs}.

\begin{table}[t]
\centering
    \begin{tabular}{lp{0.65\textwidth}p{0.20\textwidth}}
        \hline
        Sec. & Our contributions: & Main related results:  \\
        \hline
        \ref{Sec:hiders_game:lat_pans} & Algorithms for complete lattice w.r.t.~edge inclusion & 
        Thm.~\ref{The:ne_not_empt_latt_incl},~\ref{The:strong_pane_not_empt_subset_incl},\\
        & & Alg.~\ref{alg:min_incl_ne}-\ref{alg:k_min_incl_ne},\\
        & & Prop.~\ref{prop:run_time},~\ref{prop:ne_char},~\ref{prop:unil_del_enough},\\
        & & Corol.~\ref{cor:strong_pane_not_empt_subsemilatt_incl}\\
        \ref{Sec:hiders_game:gr_le_pans} & Characterisation of the greatest and the least \PANS{} & Prop.~\ref{prop:great_pane_char}, \ref{prop:least_pane_char}  \\
        \ref{Sec:hiders_game:eff_incl} & Efficiency and edge inclusion & 
        Thm.~\ref{the:mono_util_incl} \\
        \ref{Sec:hiders_game:stren_eff} & Strength and efficiency & 
        Thm.~\ref{the:max_util_iff_strong},\\ 
        & & Corol.~\ref{cor:same_util_if_strong},\\
        & & Prop.~\ref{prop:unique_util_char}\\
        \ref{Sec:hiders_game:eff_bound} & General efficiency bounds & Prop.~\ref{prop:gen_add_bound} \\
        \ref{Sec:hiders_cases:eq_alpha} & Characterisation and efficiency for equal $\alpha$'s & Thm.~\ref{the:ne_char:eq_alpha_m_0}, \ref{the:ne_char:eq_alpha_m_0:eff},\\ 
        & & Corol.~\ref{cor:ne_char:eq_alpha_m_0:str}  \\
        \ref{Sec:hiders_cases:sm_alpha_bes_1} & Characterisation and efficiency for small $\alpha$s besides $\alpha_1$& Thm.~\ref{prop:pane_char:a1_others_eq},~\ref{prop:pane_char:a1_others_eq:eff} \\
        \ref{Sec:advice:dist_player} & Policy for the players & 
        Thm.~\ref{the:mono_util_incl}, Alg.~\ref{alg:great}, 
        Prop.~\ref{prop:great_pane_char} \\
        \ref{Sec:advice:ext_obs} & Policy for the seekers & Thm.~\ref{the:mono_util_incl}, 
        Prop.~\ref{prop:great_pane_char} \\
        \hline
    \end{tabular}
    \caption{The summary of our main technical results.}
    \label{tab:our_main_res}
\end{table}

Overall, we extend the existing adversarial social network analysis by modelling
players who connect themselves and others to gain influence, while keeping a low profile as a network
formation game. 
We discover the lattice structure of the stable states and
gain important insights into their efficiency. 
We assume only the hiders being strategic, which positions 
it in step~$1$ of~\cite{MichalakRahwanWooldridge2017}, which provides several steps in modelling interactions with SNA. We are the first to consider the hiders'
interaction. 

%% file: model.tex
\section{Our Model}\label{Sec:model}
\subsectioninline{The hiders' game definition}
Consider an undirected graph $G_0 = (V, E_0)$ with nodes $V$ and edges $E_0$. As we will soon define, nodes may have different abilities to
connect, but once established, the connection is mutual. Deletion means not creating a connection. We will refer to $G_0$ as the \defined{original graph}, before the play.
Nodes $V$ are divided into 
the active nodes, called \defined{players},
$N = \set{1, \ldots, n} \subseteq V = \set{1, \ldots, n, n + 1, \ldots, n + m}$,
and inactive nodes, called \defined{non-players}, $V \setminus N$. We assume that the only connections that exist at the beginning of the game are those between non-players, i.e. $(i, j) \in E_0 \Rightarrow i, j \in V \setminus N$.\footnote{We allow the players to perform any number
of actions, thereby forming their connections, so there is no loss of generality in assuming that
the original graph connects no players.}
There are at least two players in the game, i.e.~$n \geq 2$, while $m \geq 0$.
The case when $m = 0$ models the formation of a completely new network by players~$N$, without inactive nodes, while
$m \geq 1$ stands for the players joining an existing network
of $V \setminus N$.
A \defined{strategy} $s_i$ of player $i \in N$ is
a (perhaps empty) set of the following actions:
\begin{itemize}
	\item	Initiating a connection of oneself to another
	node $j \in V$;
	this action is denoted by $\cnct(i, j)$ or merely $\cnct(j)$ if $i$ is obvious from the context.
	
	\item Connecting any non-players $j$ and $k$. This action
	is effective only if $j$ and $k$ neighbour the connecting
	player $i$; we denote this as $\cnct_i(j, k)$
	or $\cnct(j, k)$ if $i$ is clear from the context. 

\end{itemize}
The set of all strategies of player~$i$ is
$S_i \defas 2^{\set{\cnct(i, j) : j \in V \setminus \set{i}} \cup \set{\cnct(j, k) : j, k \in V \setminus N, j \neq k}}$.
%

Given the original graph $G_0 = (V, E_0)$, $E_0 \subseteq (V \setminus N) \times (V \setminus N)$, and a strategy profile $s = s_1, \ldots, s_n$, we define
the \defined{added edges} $A(s) \defas \{ (i, j) \in V \times V : $ 
\begin{eqnarray}
\begin{array}{ll}
 & \left(\left(\cnct(i, j) \in s_i\right) \land (\cnct(j, i) \in s_j) \land (i, j \in N)\right) \\
\lor &  \left((\cnct(i, j) \in s_i) \land (i \in N, j \in V \setminus N)\right) \\
\lor & \left((\cnct_k(i, j), \cnct(k, i), \cnct(k, j) \in s_k) 
\land (k \in N) \land (i, j \in V \setminus N\right) 
 ) \}.
\end{array}
\end{eqnarray}
The edges in $A(s)$ can
be either two players who wish to connect to one another,
a player wishing to connect to a non-player node, or
a player connecting two non-player neighbours of hers. 
Let the final set of edges be $E(s)\defas E_0 \cup A(s)$.

Inspired by the idea of utilities seen as payment minus
link costs~\cite{GoyalJoshi2006,BalaGoyal2000,KonigBattistonNapoletanoSchweitzer2011}, and mostly by indirect communication minus
communication costs~\cite{JacksonWolinsky1996}, we
define player~$i$'s \defined{utility} in
the \defined{resulting graph}
$G(s) = G(E(s)) \defas (V, E(s))$
as
\begin{equation}
u_i(s) = u_i(G(s)) = u_i(E(s)) \defas  \sum_{j \in \Neighb(i)}{\deg(j)} - \alpha_i \deg(i),
\label{eq:util}
\end{equation}
where $\Neighb(i) \defas \set{j : (i, j) \in E(s), j \neq i}$ is $i$'s neighbourhood, and $\deg_G(i)$ is $i$'s degree in $G$\footnote{The graph in the subscript can be omitted when it is clear from the context.}. 
The utility expresses the gain from influence minus
the damage of visibility. A player's influence is expressed by
the sum of the degree centralities of her neighbours, assuming different people have different views, so even if I hear about the same person from multiple mutual neighbours, they all contribute. A player
becomes visible because of her own degree centrality~\cite{Freeman1978}, and each connection costs her $\alpha_i$. The player-specific $\alpha_i$
denotes her cost of publicity. The real number $\alpha_i$ could even be negative, if a player likes being connected, though we assume nonnegativity as a rule, since a player typically aims to hide.
The larger $\alpha_i$ is, the less player~$i$ will be inclined to connect herself, trading off connectedness for low visibility.

There
exist infinitely many possible centrality measures, and we have chosen the simple fundamental degree centrality
to concentrate on the properties
of this new complex model. Degree centrality is
employed by many social network analysis tools~\cite{gephi2020,nodexl2020,ucinet2020} and is considered
to indicate fame~\cite[Page~$45$]{TsvetovatKouznetsov2011},
prestige~\cite[\sectn{2.2.1}]{TabassumPereiraFernandesGama2018},
and ``hubbiness''~\cite[\sectn{2.2}]{BerlingerioCosciaGiannottiMonrealePedreschi2011}. 
We scale the degree centrality by a player-specific coefficient
$\alpha_i \geq 0$, expressing the damage of own visibility.
\anote{Another way to define utility is by the rank of centrality.}
This completes the definition of the \defined{hiders' game}.

This is a game of \textbf{strategic complements w.r.t.~the actually newly added
edges}, i.e.~$\forall i \in N, E \subset E', F \cap E' = \emptyset$, $$u_i(E' \cup F) - u_i(E') \geq u_i(E \cup F) - u_i(E).$$

Connecting with others increases the sum of the degrees of the neighbours, while increasing one's own degree centrality, but connecting non-player neighbours costs
nothing. Any edge creation (between a player and a player, a player and a
non-player or two non-players) could be explicitly charged for, but charging a player for connecting herself to another player of
to a non-player can be incorporated in her coefficient $\alpha_i$
multiplying her own degree. This is not the case for charging a player for connecting
two neighbouring non-players, but charging for that is problematic in
the first place, as it is often unclear, which neighbouring player to charge,
since given the static approach, the identity of the neighbour connecting
two non-players may be unclear. Therefore, such a charging should be only modelled 
in dynamic models. 

We remark that the existence of non-player nodes is \emph{not} equivalent to merely adding a constant amount of utility to each player, 
since player~$i$ will not connect to a non-player, if that non-player's
degree is less than $\alpha_i$. 

%
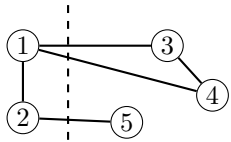
\begin{SCfigure}[1.2]
\begin{tikzpicture}[scale=0.6]
\tikzstyle{vertex} = [circle, draw=black, minimum size=12pt,inner sep=0pt]
\tikzstyle{title} = [circle, draw=white, minimum size=25pt,inner sep=0pt, font=\large\bfseries]
\tikzstyle{edge} = [-,thick]

\draw [dashed, thick, black] (0,1.3) -- (0,-1.8);

\node[vertex](v1) at (-1, 0.4){1};
\node[vertex](v2) at (-1, -1.2){2};



\node[vertex](vn1) at (2.2, 0.4){3};
\node[vertex](vn2) at (3.2, -0.7){4};


\node[vertex](vnm) at (1.3, -1.3){5};

\draw[edge] (v1) -- (v2);
\draw[edge] (v1) -- (vn1);
\draw[edge] (v1) -- (vn2);
\draw[edge] (vn1) -- (vn2);
\draw[edge] (v2) -- (vnm);

\end{tikzpicture}
\caption{Players~$N = \set{1,2}$ and non-players~$V \setminus N = \{3,4,5\}$.
}%
\label{fig:hid_game}\vspace{-0.3cm}
\end{SCfigure}

We illustrate the above definition with the following example.
\begin{example}[Industrial intelligence]\label{ex:ind_espion}
Companies $A$ and $B$ try to infiltrate company $C$. Company $A$'s mole
is denoted by~$1$,
and $B$'s mole by~$2$.
Initially, company $C$ consists of three employees
$\set{3, 4, 5}$. Hence, the set of players is $N = \set{1, 2}$, while the set of non-players is $\set{3, 4, 5}$, as depicted in \figref{fig:hid_game}. Let $\alpha_1 = 1$ and $\alpha_2 = 0.5$, express that mole $1$ fears suspicion twice as much as $2$ does.

We now consider the depicted situation, which we later see is not an \NE.
Firstly, in our model, an edge between two players is added if and only if both make that
connection. In \figref{fig:hid_game}, player~$1$ initiates the connection to~$2$ and $2$ to~$1$,
so they are interconnected. Secondly, a player is always connected to the non-players to whom the player initiates connection. In \figref{fig:hid_game}, player~$1$ connects to non-players~$3$ and~$4$, while $2$ to $5$. Thirdly, any player can connect two of its non-player neighbours. Here, player~$1$ connects $3$ and $4$. For clarity, in this simple example we assume that the non-players all start disconnected, i.e., that the original graph is $G_0 = \{\{3,4,5\},\emptyset\}$. 
In $G(s)$, player $1$'s utility is equal to
$u_1(s) = \deg(2) + \deg(3) + \deg(4) - \alpha_1 \deg(1) = 3$,
while $u_2(s) = \deg(1) + \deg(5) - \alpha_2 \deg(2) = 3$. 
\end{example}

In summary, our model generalises existing models (see
\sectnref{Sec:related_work}) by having non-player nodes
and letting any player to connect her non-player neighbours. 

\subsectioninline{Solution concepts}
As concluded by Jackson~\cite{Jackson2003}, the standard solutions like
Nash equilibria or \SPE{} are usually inapplicable for networks, the
former being  too permissive, such as allowing the edgeless graphs to constitute an equilibrium regardless of the parameters, when both nodes do not connect, while the
latter being too detail-sensitive. Therefore, we use the concept of the
pairwise Nash stability~\cite{BlochJackson2006}, building upon the classical concept of \NE{},
while avoiding its excessive permissiveness. In particular, it requires any two unconnected
players who would benefit from interconnecting to be connected.

We need to define, for profile $s \in S$ and players $i, j \in N$,
the profile that adds edge $(i, j)$.
Formally, we denote by $s + (i, j)$ the profile (w.l.o.g., $i < j$)
$(s_1, \ldots, s_{i - 1}, s_i \cup \set{\cnct(i, j)}, s_{i + 1}, \ldots, s_{j - 1}, s_j \cup \set{\cnct(j, i)}, s_{j + 1}, \ldots, s_n)$.
\begin{defin}\label{def:pairwise_nash}
Profile $s \in S \defas S_1 \times \ldots \times S_n$ is a
\defined{$k$-\text{strong} Nash equilibrium, denoted $k$-\NE}%
~\cite{BlochJackson2006} if it is a $k$-\text{strong} \NE{} of the hiders' game.
Formally, for every $k$-coalition $P \subseteq N, \abs{P} \leq k, s'_P \in S_P \defas \times_{i \in P}{S_i}$, if
$\exists i \in P$, such that $u_i(s'_P, s_{N \setminus P}) > u_i(s_N)
\Rightarrow \exists j \in P$, such that $u_j(s'_P, s_{N \setminus P}) < u_j(s_N)$.

Next, 
$s \in S$ is a \defined{pairwise $k$-\text{strong} Nash equilibrium, denoted $k$-\PANE},
if the resulting graph is also pairwise stable~\cite{JacksonWolinsky1996}, i.e.~%
$\forall (i, j) \in \set{(r, l) : r, l \in N} \setminus A(s) : 
 u_i(s) < u_i(s+(i, j)) \Rightarrow u_j(s) > u_j(s+(i, j))$. 
%
Slightly abusing notation, we denote the set of all the
$k$-\text{strong} \NE{} as $k$-\NE,
and the set of all the pairwise
$k$-\text{strong} \NE{} as $k$-\PANE. By definition, $k-\PANE \subseteq k-\NE$.
\end{defin}
Let us simplify the matters.
\begin{observation}
When talking about $k$-\text{strong} \PANE{} for $k \geq 2$, the pairwise
stability requirement becomes excessive, because any coalition of $k$ players is able to connect if they all agree.
\end{observation}

Next, we define the analogous notions for graphs.
\begin{defin}
A graph is \defined{$k$-\text{strong} Nash stable or
$k$-\NS} if it can result from a $k$-\text{strong} \NE. 
It is called \defined{pairwise $k$-\text{strong} Nash stable or
$k$-\PANS} if it can result from a pairwise $k$-\text{strong} \NE. Abusing
notation, we also denote the the set of all such graphs as $k$-\NS{} or $k$-\PANS{}, respectively. Now, by definition, $k-PANS \subseteq k-\NS$.
\end{defin}
A $1$-strong Nash equilibrium is just called
a Nash equilibrium, and an $n$-strong equilibrium is just called a strong equilibrium. We name pairwise equilibria and stable graphs analogically.

The profile in Example~\ref{ex:ind_espion} is not an equilibrium,
since the only Nash equilibrium there is a completely connected graph.
An example with multiple possible equilibria appears in
\figref{fig:subsemilatt_non_mono}.

By definition, the utilities of the players depend only on the
resulting graph, and so 
we would like to identify the resulting graph with the set of the strategy profiles
that result in this graph. However, a profile where an edge
is never added is not identical to a profile where one player wants to add this
edge while the other player does not. Moreover, the resulting graph
is not even sufficient to decide whether the strategy profile is a
pairwise \NE, as the following example demonstrates.
\begin{example}\label{example:res_graph_pane_dep_strat}
Consider players $N = \set{1, 2}$ with $\alpha_1 = 0.1$ and $\alpha_2 = 2$.
The resulting empty graph is an
equilibrium if $s_1 = \emptyset$ and $s_2 = \emptyset$,
since no player can unilaterally change the resulting graph
(it is even a \PANE, since player~$2$ prefers to stay separate),
but not an equilibrium if $s_2 = \set{\cnct(2, 1)}$, since in the latter case,
player $1$ can connect by deviating to $s'_1 = \set{\cnct(1, 2)}$, and 
$u_1(s'_1, s_2) > u_1(s_1, s_2)$. 
%
Therefore, the empty graph here is a \PANS{}, because it can result from a \PANE{}, but it does not have to result from an equilibrium.
\end{example}

In order to be able to identify strategy profiles with the resulting
graphs, we assume, unless said otherwise, that 
every player plays the inclusion-wise minimal
strategy resulting in the current graph, given
the strategies of the other players.
Thus, either both player connect to one another,
or both do not.
%
Now, given any graph~$G = (V, E)$ 
and $s, s' \in S$, $E(s) = E(s')$ implies
$s = s'$.
We observe
\begin{observation}
If a given graph can result from a (pairwise) $k$-strong \NE{} $s$, 
then it also results from a minimal strategy $t$ that is a (pairwise) $k$-strong \NE{}.
\end{observation}
\begin{proof}
The resulting graph of any $k$-deviation from $t$ can be attained with an at most $k$-deviation from $s$, therefore the non-existence of a $k$-deviation from $s$ implies no such deviation from $t$ exists.
\end{proof}

When analysing efficiency, 
consider the
\defined{social welfare}, defined as~
$\sw(s) = \sw(G(s)) = \sw(E(s)) \defas \sum_{i \in N}{u_i(s)}$.
%
We adapt the efficiency measures of Nash equilibria,
namely the price of anarchy%
~\cite{KP99,KoutsoupiasPapadimitriou2009,Papadimitriou2001} and
the price of stability~\cite{SchulzStierMoses2003,AnshelevichDasGuptaKleinbergTardosWexlerRoughgarden04},
to \PANE{} as follows:
$$\poa \defas \frac{\max_{s \in S}{\sw(s)}}{\min_{s \in \PANE}{\sw(s)}},
\pos \defas \frac{\max_{s \in S}{\sw(s)}}{\max_{s \in \PANE}{\sw(s)}},
$$
where $0 / 0$ is defined to be $1$, since in that case, the equilibrium has still the same
efficiency as the optimum. The $k$-\text{strong} cases are treated
analogically and denoted $k-\poa$ and $k-\pos$, respectively.
In Example~\ref{ex:ind_espion}, the only Nash equilibrium
is completely connected, achieving the maximum social welfare,
thus $\poa = \pos = 1$.

%% file: hiders_game.tex
\section{Game Lattice and Efficiency}\label{Sec:hiders_game}

In this section, we algorithmically prove the existence of pairwise Nash equilibria and analyse their structure and efficiency. 

\subsectioninline{Lattice of pairwise Nash stable graphs}
\label{Sec:hiders_game:lat_pans}
We first observe that the strategy set $S_i$, with the partial order of set inclusion,
constitutes a complete lattice. Indeed, for any $E_i \subseteq S_i$,
$\sup(E_i) = \cup_{s \in E_i}{s}, \inf(E_i) = \cap_{s \in E_i}{s}$.
We will also prove that the equilibria constitute a lattice with respect
to edge inclusion. We will prove this algorithmically, because the other
natural proofs do not work, as we show in \sectnref{}.
For the sake of this, we will need the following definition:
\begin{defin}
For any non-player $i \not \in N$, define \defined{$i$'s effective degree for $k \in N$},
denoted as $\deg(i; k)$,
as $|\{j \in V \setminus N : \cnct(k, i), \cnct(k, j), \cnct_k(i, j) \in s_k,
\nexists l \in N, \cnct(l, i), \cnct(l, j), \cnct_l(i, j) \in s_l\}|$.
\end{defin}
Namely, $\deg(i; k)$ is the number of $k$'s neighbours he uniquely connects to $i$. 
Now, we present algorithms for finding \PANE.

\RestyleAlgo{ruled}

\begin{algorithm}[t!]
\caption{\textbf{MinIncluding${\PANS}$} $\paren{(V, E), E_0, (\alpha_i)_{i = 1}^{n + m}}$}\label{alg:min_incl_ne}
\SetAlgoVlined
\LinesNumbered


\vspace{0.05cm}
	$c \leftarrow True$ ;
	\tcp{While we can improve:}
	\While {$c = True$}{
	    \label{alg:min_incl_ne:imp}
	    $c \leftarrow False$;\\
	    \For {$i \in N$}{
	    \label{alg:min_incl_ne:uni_imp}
	    \tcp{Interconnect $i$'s non-player neighbours:}
	    \For {$j, k \not \in N$, s.t.~$j \neq k, (k, j) \not \in E, (i, j), (i, k) \in E$}{
	    \label{alg:min_incl_ne:uni_imp:fromS}
	        $E \leftarrow E \cup \bigcup\set{(j, k)}$;\ \\
	        $c \leftarrow True$;\
	    }
	    \tcp{Connect to profit.\ non-play. neighbors:}
	    \While {$\exists T \subset V \setminus N$, s.t.\ 
    		adding $\set{(i, j)}_{j \in T}$ to $E$, and then 
    		adding $\set{(k, l)}_{k \neq l, k, l \in N(i) \cap (V \setminus N)}$ to $E$
    		strictly increases $u_i(E)$,}{
    		\label{alg:min_incl_ne:uni_imp:fromS:connect_to_nonplayer}
    		    Take any inclusion-wise minimum such $T$\\
    		    $E \leftarrow E \cup \set{(i, j)}_{j \in T} \cup \set{(k, l)}_{k \neq l, k, l \in N(i) \cap (V \setminus N)}$; \\
    		    $c \leftarrow True$;\ \\
		}
	    }
	    
	    \tcp{Maintain pairwise stability:}
	    \For {each $i, j \in N$}{
	    	\label{alg:min_incl_ne:uni_imp:edge_add}
        	\If {adding $(i, j)$ to $E$, would imply ($\deg(i) > \alpha_j$ and $\deg(j) \geq \alpha_i$)
        	or ($\deg(i) \geq \alpha_j$ and $\deg(j) > \alpha_i$)}{
        	    $E \leftarrow E \cup \set{(i, j)}$;\ \\
        	    $c \leftarrow True$;\
        	}
    	}
	}

\Return{the (updated) graph~$(V, E)$;}

\end{algorithm}

\begin{algorithm}[t!]
\caption{\textbf{MaxIncluded${\PANS}$} $\paren{(V, E), E_0, (\alpha_i)_{i = 1}^{n + m}}$}\label{alg:max_incled_ne}
\SetAlgoVlined
\LinesNumbered


\vspace{0.05cm}
    $c \leftarrow True$ ;
	\tcp{While we can improve:}
	\While {$c = True$}{
	$c \leftarrow False$
	
	\tcp{While some player can profitably delete, do it:}
	\label{alg:max_incled_ne:uni_imp}
	\While {$\exists i, j \in N, j \not = i$, such that 
            $(i, j) \in E$ and
    		$\deg(j) < \alpha_i$}{ 
    	$E \leftarrow E \setminus \set{(i, j)}$;\ \\
    	$c \leftarrow True$ ;
	}
	\While {$\exists i \in N, j \in V \setminus N$, such that 
            $(i, j) \in E$ and
    		$\deg(j) + \deg(j; i) < \alpha_i$}{ 
    	$E \leftarrow E \setminus \set{(i, j)}$;\ \\
    	$c \leftarrow True$ ;
	}

	\tcp{clean up the edges between non-players}
	\While{$\exists (i, j) \in E \setminus E_0$, s.t.~%
	    $\set{k \in N : (k, i), (k, j) \in E} = \emptyset$}{
	        $E \leftarrow E \setminus \set{(i, j)}$;\ \\
	        $c \leftarrow True$ ;
	}
	}

\Return{the (updated) graph~$(V, E)$;}

\end{algorithm}


%

\begin{algorithm}[t!]
\caption{\textbf{Least${\PANS}$}$\paren{G = (V, E), E_0, (\alpha_i)_{i = 1}^{n + m}}$}\label{alg:least}
\SetAlgoVlined
\LinesNumbered


\vspace{0.05cm}

\Return{\textbf{MinIncluding${\PANS}$}$\paren{G_0 \defas (V, E_0), E_0, (\alpha_i)_{i = 1}^{n + m}}$;}

\end{algorithm}

%

\begin{algorithm}[t!]
\caption{\textbf{$\sup{\PANS}$}$\paren{G = (V, E), E_0, (\alpha_i)_{i = 1}^{n + m}, s, t}$}\label{alg:sup}
\SetAlgoVlined
\LinesNumbered


\vspace{0.05cm}

$G' \defas (V, E(s) \cup E(t))$;\ \label{alg:sup:init_unite}\\

\Return{\textbf{MinIncluding${\PANS}$}$\paren{G', E_0, (\alpha_i)_{i = 1}^{n + m}}$;}

\end{algorithm}

%

\begin{algorithm}[t!]
\caption{\textbf{Greatest${\PANS}$}$\paren{G = (V, E), E_0, (\alpha_i)_{i = 1}^{n + m}}$}\label{alg:great}
\SetAlgoVlined
\LinesNumbered


\vspace{0.05cm}

\label{alg:great:init_all}
$G' \defas (V, \set{(i, j) : i, j \in V, i \neq j})$;\
\tcp{the complete graph}

\Return{\textbf{MaxIncluded${\PANS}$}$\paren{G', E_0, (\alpha_i)_{i = 1}^{n + m}}$;}

\end{algorithm}

%

\begin{algorithm}[t!]
\caption{\textbf{$\inf{\PANS}$}$\paren{G = (V, E), E_0, (\alpha_i)_{i = 1}^{n + m}, s, t}$}\label{alg:inf}
\SetAlgoVlined
\LinesNumbered


\vspace{0.05cm}

\label{alg:sup:init_inter}
$G' \defas (V, E(s) \cap E(t))$;\ \label{alg:inf:init_unite}\\

\Return{\textbf{MaxIncluded${\PANS}$}$\paren{G', E_0, (\alpha_i)_{i = 1}^{n + m}}$;}

\end{algorithm}
%
\begin{theorem}\label{The:ne_not_empt_latt_incl}
The set of all the pairwise Nash stable graphs constitutes a non-empty lattice with respect
to the inclusion order of the edge sets, where the least
element is given by Algorithm~\ref{alg:least} and
the greatest possible pairwise Nash stable graph is returned by
Algorithm~\ref{alg:great}. The
supremum (join) of two \PANS{} is attained by Algorithm~\ref{alg:sup},
and their infimum (meet)
is provided by Algorithm~\ref{alg:inf}.
\end{theorem}
In short, Algorithm~\ref{alg:min_incl_ne} makes unilateral connections by players to a non-player or between two neighbouring non-players, as long as they benefit the initiating player. It also interconnects pairs of players, such that no member of a pair loses and at least one strictly benefits from it. Similarly, Algorithm~\ref{alg:max_incled_ne} severs connections unilaterally, while is profits the severing player.

In particular, Algorithm~\ref{alg:sup} minimally extends two pairwise Nash stable networks to a pairwise stable
network containing all the original connections. 
Loosely speaking,
the greedy approach is optimum because 
of the complementarity of edge addition.
As for the running time, the following holds: 
\begin{proposition}\label{prop:run_time}
Algorithms~\ref{alg:min_incl_ne}, \ref{alg:least}, and
\ref{alg:sup} take time $O((n + m)^3 n (n + m^2 \cdot 2^m))$, while
Algorithms~\ref{alg:max_incled_ne}, \ref{alg:great}, and
\ref{alg:inf} take $O((n + m)^3 n (n + m^2 \cdot 2^m))$. 
\end{proposition}

We will later provide several important characterisations,
alleviating the need to face this complexity practically.
%
%

Let us first extend the lattice result with a characterisation of a \PANE{}. This is basically an equivalent definition, explaining the notion of PANE. 

\begin{proposition}\label{prop:ne_char}
A minimal profile $s$, resulting in graph $G(s) = (V, E(s))$, is a \PANE{} if and only if all the following conditions hold:
\begin{enumerate}
	\item \label{prop:ne_char:playnoplay}
	For any~$i \in N$, 
	\begin{itemize}
	    \item   For any~$j \not \in N$, $(i, j) \in E(s)$ 
	    $\Rightarrow \deg(j) + \deg(j; i) \geq \alpha_i$, 
	    \item   
	    there exists no $T \subset V \setminus N$, s.t.\ 
    		adding $\set{(i, j)}_{j \in T}$ to $E$, and then 
    		adding $\set{(k, l)}_{k \neq l, k, l \in N(i) \cap (V \setminus N)}$ to $E$
    		strictly increases $u_i(E)$;
	    {\bf and}
	\end{itemize}
	
	\item   \label{prop:ne_char:playplay}	
	For any~$i, j \in N, i \neq j$,
	\begin{itemize}
	    \item   $(i, j) \in E(s)$ 
	    $\Rightarrow \deg(j) \geq \alpha_i, \deg(i) \geq \alpha_j$, 
	    \item   
        $\deg_{E(s) \cup \set{(i, j)}}(i) \geq \alpha_j, 
	    \deg_{E(s) \cup \set{(i, j)}}(j) \geq \alpha_i, 
	    \deg_{E(s) \cup \set{(i, j)}}(i) + \deg_{E(s) \cup \set{(i, j)}}(j) > \alpha_j + \alpha_i \Rightarrow (i, j) \in E(s)$; {\bf and}
	\end{itemize}
	
	\item   \label{prop:ne_char:noplaynoplay}
	$\exists i \in N, j, k \in V \setminus N$, 
	$(i, j), (i, k) \in E$ $\Rightarrow$
	$(j, k) \in E$.
\end{enumerate}
\end{proposition}
We now generalise the lattice Theorem~\ref{The:ne_not_empt_latt_incl}
to $k$-strong \PANE.
\begin{theorem}\label{The:strong_pane_not_empt_subset_incl}
The $k$-strong pairwise stable graphs form a non-empty lattice
with respect to the edge-inclusion order.
Algorithm~\ref{alg:min_incl_ne} is generalised
by Algorithm~\ref{alg:k_min_incl_ne},
Algorithm~\ref{alg:max_incled_ne}
is generalised analogously, and the other algorithms remain unchanged.
Moreover, the $k + 1$-strong pairwise stable graphs form a
subset\footnote{It is a subset, but not necessarily a sublattice,
since the actions of infimum and supremum may differ between the lattices.}
of all the $k$-strong pairwise stable graphs,
for each $k \geq 1$.
\end{theorem}
Algorithm~\ref{alg:k_min_incl_ne} makes connections on the part of coalitions of size up to~$k$, as long as nobody in such a coalition loses, and at least one player there strictly benefits, generalising Algorithm~\ref{alg:min_incl_ne}.

\begin{algorithm}[t!]
\caption{\mbox{\textbf{MinIncluding${k-\PANS}$} $\paren{(V, E), E_0, (\alpha_i)_{i = 1}^{n + m}}$}}\label{alg:k_min_incl_ne}
\SetAlgoVlined
\LinesNumbered


\vspace{0.05cm}
	$c \leftarrow True$ ;
	\tcp{While we can improve:}
	\While {$c = True$}{
	    \label{alg:k_min_incl_ne:uni_imp}
	    $c \leftarrow False$;\\
	    \For {$i \in N$}{
	    \label{alg:k_min_incl_ne:uni_imp:fromS}
	    \tcp{Interconnect $i$'s non-player neighbours:}
	    \For {$j, k \not \in N$, s.t.~$j \neq k, (k, j) \not \in E, (i, j), (i, k) \in E$}{
	        \label{alg:k_min_incl_ne:uni_imp:fromS:non_pl_neigh}
	        $E \leftarrow E \cup \bigcup\set{(j, k)}$;\ \\
	        $c \leftarrow True$;\
	    }
	    }

	    \tcp{Coalitions connect to profit.\ neighbours:}
        \While {$\exists U, W \subseteq N, \abs{U \cup W} \leq k$, such that
        {
	    	\begin{itemize}
        	\item[(1)] $\exists T \subseteq V \setminus N$,
        	such that adding $\set{(i, j) : i \in U, j \in T}$, \\and
        	then interconnecting all the nodes in
        	$\bigcup{N(i)}_{i \in U} \cap (V \setminus N)$, 
        	\item[(2)]	\label{alg:min_incl_ne:uni_imp:fromS:coal_connect:players}
        	and there exist
        	$\set{i_l, j_l}_{l = 1}^{d}  \subset W, (i_l, j_l) \not \in E, \forall l = 1, \ldots, d$ \\such that
        	adding $\set{(i_l, j_l)_{l = 1}^d}$, 
        	\item[(3)] results in $E'$, such that $\forall i \in U \cup W, u_i(E') \geq u_i(E)$
        	and $\exists j \in U \cup W$, such that $u_j(E') > u_j(E)$,
        	\end{itemize}
        }
        }{
        \label{alg:min_incl_ne:uni_imp:fromS:coal_connect}
            {
            Take any inclusion-wise minimum such $U, W$ and $T$,
            in the sense that no set(s) can be made strictly smaller without changing the others
            
            \label{}
        	$E \leftarrow E \cup \set{(i, j)}_{j \in T, i \in U} \cup \bigcup\set{(k, l)}_{k \neq l, k, l \in \bigcup{N(i)}_{i \in U} \cap (V \setminus N)}
        	\cup \set{(i_l, j_l)}_{l = 1}^d$;\ \\
        	
        	$c \leftarrow True$;\
        	}
		}
	}

\Return{the (updated) graph~$(V, E)$;}

\end{algorithm}

Remarkably, we do not need to adapt Algorithm~\ref{alg:max_incled_ne} to
the $k$-strong case, since unlike
addition, where one added edge can make another addition profitable,
a deletion is either profitable or not. Together with 
Lemma~\ref{lemma:alg:del_one_enough} from~\ref{Sec:omit_proof},
we conclude that if a set of players can benefit from some deletions, then single player can benefit from merely one deletion, too.
Formally:
\begin{proposition}\label{prop:unil_del_enough}
If $\exists S \subseteq N, F \subseteq \cup_{i \in S}{\set{\set{i} \times \Neighb(i)}}$, such that
$\forall i \in S: u_i(E \setminus F) \geq u_i(E)$ and $\exists i \in S,
u_i(E \setminus F) > u_i(E)$,
i.e.~a set deletion is profitable, then
$\exists H \subseteq \set{i} \times \Neighb(i)$, such that $u_i(E \setminus H) > u_i(E)$, so a unilateral deletion of player~$i$ strictly increases $u_i$.
%
\end{proposition}

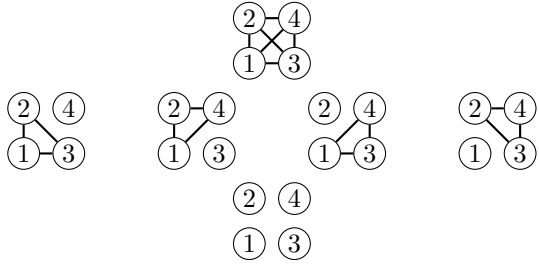
\begin{figure}[t]
\center

\begin{tikzpicture}

\tikzstyle{vertex} = [circle, draw=black, minimum size=12pt,inner sep=0pt]
\tikzstyle{title} = [circle, draw=white, minimum size=25pt,inner sep=0pt, font=\large\bfseries]
\tikzstyle{edge} = [-,thick]

\begin{scope}[shift={(3,4.2)}]


\node[vertex](v1) at (0, 0){1};
\node[vertex](v2) at (0, 0.6){2};
\node[vertex](v3) at (0.6, 0){3};
\node[vertex](v4) at (0.6, 0.6){4};

\draw[edge] (v1) -- (v2);
\draw[edge] (v1) -- (v3);
\draw[edge] (v1) -- (v4);
\draw[edge] (v2) -- (v3);
\draw[edge] (v2) -- (v4);
\draw[edge] (v3) -- (v4);

\end{scope}

\begin{scope}[shift={(0,3)}]

\node[vertex](v1) at (0, 0){1};
\node[vertex](v2) at (0, 0.6){2};
\node[vertex](v3) at (0.6, 0){3};
\node[vertex](v4) at (0.6, 0.6){4};

\draw[edge] (v1) -- (v2);
\draw[edge] (v2) -- (v3);
\draw[edge] (v3) -- (v1);

\end{scope}

\begin{scope}[shift={(2,3)}]

\node[vertex](v1) at (0, 0){1};
\node[vertex](v2) at (0, 0.6){2};
\node[vertex](v3) at (0.6, 0){3};
\node[vertex](v4) at (0.6, 0.6){4};

\draw[edge] (v1) -- (v2);
\draw[edge] (v2) -- (v4);
\draw[edge] (v4) -- (v1);

\end{scope}

\begin{scope}[shift={(4,3)}]

\node[vertex](v1) at (0, 0){1};
\node[vertex](v2) at (0, 0.6){2};
\node[vertex](v3) at (0.6, 0){3};
\node[vertex](v4) at (0.6, 0.6){4};

\draw[edge] (v1) -- (v3);
\draw[edge] (v3) -- (v4);
\draw[edge] (v4) -- (v1);

\end{scope}

\begin{scope}[shift={(6,3)}]

\node[vertex](v1) at (0, 0){1};
\node[vertex](v2) at (0, 0.6){2};
\node[vertex](v3) at (0.6, 0){3};
\node[vertex](v4) at (0.6, 0.6){4};

\draw[edge] (v2) -- (v3);
\draw[edge] (v3) -- (v4);
\draw[edge] (v4) -- (v2);

\end{scope}

\begin{scope}[shift={(3,1.8)}]

\node[vertex](v1) at (0, 0){1};
\node[vertex](v2) at (0, 0.6){2};
\node[vertex](v3) at (0.6, 0){3};
\node[vertex](v4) at (0.6, 0.6){4};

\end{scope}

\end{tikzpicture}
\caption{Assuming $\alpha_1 = \ldots = \alpha_4 = 1.5$, this is the lattice of \PANS;
a graph includes another graph if and only if it appears higher in the picture.
}%
\label{fig:example:subsemilatt_3}
\end{figure}

\begin{proof}
Fix profile $s$.
We prove that if there is a profitable set of deletions of players in $S$
such that the $u_j$s of all $j \in S$ do not decrease and for at least one
$i \in S$, the $u_i$ strictly increases, then there exists a single player in
$S$ who
can profit from a unilateral deletion.
We express the result of deletions as the new profile $s'_S \subsetneq s_S$,
and assume that $\forall i \in S$,
$u_j(s'_S, s_{N \setminus S}) \geq u_j(s)$, while $\exists i \in S$,
such that $u_i(s'_S, s_{N \setminus S}) > u_i(s)$.
Undoing all the deletions that do not touch $i$ cannot
decrease $i$'s utility. We can achieve that effect from $s$ by solely a change to $i$'s strategy, namely taking $s'_i$ and additionally deleting all the connections
the have been deleted by the other side, namely
we define $s''_{i} \defas s'_i \setminus
\set{\cnct(i, j) : \cnct(j, i) \in s_j \setminus s'_j, j \in S \setminus \set{i}}$. Hence, we conclude that
$u_i(s''_{i}, s_{N \setminus \set{i}}) \geq u_i(s'_S, s_{N \setminus S}) > u_i(s)$,
so $i$ can profit by her unilateral deletions as well.
\end{proof}
Proposition~\ref{prop:unil_del_enough} means that
the join action is the same in the $k-\PANS$ lattices from
Theorem~\ref{The:strong_pane_not_empt_subset_incl} for any $k$; formally,
we conclude the following.
\begin{corollary}\label{cor:strong_pane_not_empt_subsemilatt_incl}
For each $k \geq 1$, the join semi-lattice of $k + 1$ strong pairwise-stable graphs is
a sub semi-lattice of the join semi-lattice of $k$-strong pairwise-stable graphs.
\end{corollary}

Together with Lemma~\ref{lemma:alg:del_one_enough} from
\sectnref{Sec:omit_proof}, we conclude
\begin{proposition}\label{prop:unil_del_one_enough}
If $\exists S \subseteq N, F \subseteq \cup_{i \in S}{\set{\set{i} \times \Neighb(i)}}$, such that
$\forall i \in S: u_i(E \setminus F) \geq u_i(E)$ and $\exists i \in S,
u_i(E \setminus F) > u_i(E)$,
i.e.~a set deletion is profitable, then
$\exists (i, j) \in \set{i} \times \Neighb(i)$, such that $u_i(E \setminus \set{(i, j)}) > u_i(E)$, so a single deletion of player~$i$ strictly increases $u_i$.
%
\end{proposition}

We now demonstrate the lattice of pairwise Nash stable graphs and
its sub semi-lattices of $k$-strong \PANS{} for various $k$s on a concrete example.
\begin{example}\label{example:subsemilatt_3}
Let the original graph $G$ be the empty graph on $4$
players $N = \set{1, 2, 3, 4} = V$, and 
assume $\alpha_1 = \ldots = \alpha_4 = 1.5$. 
\figref{fig:example:subsemilatt_3} depicts the lattice
of pairwise Nash stable graphs, which coincides here with
the lattice of $2$-strong \PANS.
The lattices of $3$- and 4-strong pairwise stable graphs
consist only from the clique on $N$. 
\end{example}

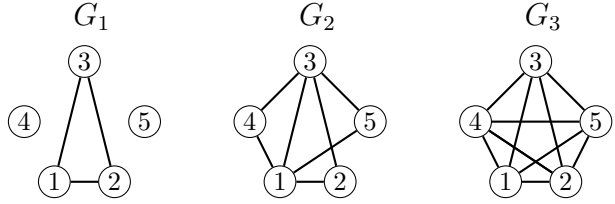
\begin{figure}[t]
\center

\begin{tikzpicture}

\tikzstyle{vertex} = [circle, draw=black, minimum size=12pt,inner sep=0pt]
\tikzstyle{title} = [circle, draw=white, minimum size=25pt,inner sep=0pt, font=\large\bfseries]
\tikzstyle{edge} = [-,thick]

\node[title] at (0.5, 2.2) {$G_1$};

\node[vertex](v1) at (0, 0){1};
\node[vertex](v2) at (0.8, 0){2};
\node[vertex](v3) at (0.4, 1.6){3};
\node[vertex](v4) at (-0.4, 0.8){4};
\node[vertex](v5) at (1.2, 0.8){5};

\draw[edge] (v1) -- (v2);
\draw[edge] (v1) -- (v3);
\draw[edge] (v2) -- (v3);

\begin{scope}[shift={(3,0)}]

\node[title] at (0.5, 2.2) {$G_2$};

\node[vertex](v1) at (0, 0){1};
\node[vertex](v2) at (0.8, 0){2};
\node[vertex](v3) at (0.4, 1.6){3};
\node[vertex](v4) at (-0.4, 0.8){4};
\node[vertex](v5) at (1.2, 0.8){5};

\draw[edge] (v1) -- (v2);
\draw[edge] (v1) -- (v3);
\draw[edge] (v2) -- (v3);
\draw[edge] (v3) -- (v4);
\draw[edge] (v3) -- (v5);
\draw[edge] (v1) -- (v4);
\draw[edge] (v1) -- (v5);

\end{scope}

\begin{scope}[shift={(6,0)}]

\node[title] at (0.5, 2.2) {$G_3$};

\node[vertex](v1) at (0, 0){1};
\node[vertex](v2) at (0.8, 0){2};
\node[vertex](v3) at (0.4, 1.6){3};
\node[vertex](v4) at (-0.4, 0.8){4};
\node[vertex](v5) at (1.2, 0.8){5};

\draw[edge] (v1) -- (v2);
\draw[edge] (v1) -- (v3);
\draw[edge] (v2) -- (v3);
\draw[edge] (v3) -- (v4);
\draw[edge] (v3) -- (v5);
\draw[edge] (v1) -- (v4);
\draw[edge] (v1) -- (v5);
\draw[edge] (v2) -- (v4);
\draw[edge] (v2) -- (v5);
\draw[edge] (v4) -- (v2);
\draw[edge] (v4) -- (v5);

\end{scope}

\end{tikzpicture}
\caption{Assuming $\alpha_1 = \alpha_2 = \alpha_3 = 1.1$ and $\alpha_4 = \alpha_5 = 3.1$,
these graphs are pairwise Nash stable.
}%
\label{fig:subsemilatt_non_mono}
\end{figure}

An interesting question is whether the discovered subsemi-lattices are
\defined{convex},
i.e.~whether for all $k$-strong \PANS{} $G_1, G_2, G_3$, having
$G_1$ and $G_3$ being $k+1$ strong and $G_1 \subseteq G_2 \subseteq G_3$
implies that $G_2$ is $k + 1$ strong too. Graph inclusion means that
the node sets are the same and the edge sets are included.
These subsemi-lattices are \emph{not necessarily convex}, as the following example demonstrates.
\begin{example}\label{example:subsemilatt_non_mono}
\figref{fig:subsemilatt_non_mono} depicts a graph on $5$ players,
where graph~$G_1$ is $3$-strong \PANS, $G_2$ is
$2$-strong and $G_3$ is $5$-strong. Since
$G_1 \subseteq G_2 \subseteq G_3$, this example counters the convexity
of the subsemi-lattice of the $3$-strong \PANS.

\end{example}

In Example~\ref{example:subsemilatt_non_mono}, the complete graph is a
strong \PANS. This is not a coincidence, as we shall prove in
Theorems~\ref{the:mono_util_incl} and~\ref{the:max_util_iff_strong}.


\subsectioninline{The greatest/least \PANS}
\label{Sec:hiders_game:gr_le_pans}
Aiming at structural insight and practical efficiency,
we characterise the inclusion-wise largest and the smallest \PANE.
We begin with the greatest \PANE, which is strong, since
Proposition~\ref{prop:unil_del_enough} implies that Algorithm~\ref{alg:max_incled_ne} works the same for strong \PANS.
W.l.o.g., $\alpha_1 \leq \ldots \leq \alpha_n$.
\begin{proposition}\label{prop:great_pane_char}
Define $p \defas \max\set{i \in N : i + m - 1 \geq \alpha_i}$, where
$\max\emptyset \defas 0$.
The resulting graph $G' = (V, E')$ returned by Algorithm~\ref{alg:great}
fulfills the following. If $p = 0$, then $G' = G_0$, i.e., all the players are disconnected. Otherwise, there exists a fully connected component from nodes
$C \defas \set{1, \ldots, p} \cup (V \setminus N)$,
and all the other nodes are independent, meaning that
$E' = \set{(j, l) : j, l \in C}$.
\end{proposition}
\begin{proof}
Consider the execution of Algorithm~\ref{alg:great} after it has invoked
Algorithm~\ref{alg:max_incled_ne}, which is iterating over players
till no one wants to disconnect.
We first prove by induction on Algorithm~\ref{alg:max_incled_ne} 
that all the nodes in $C$ remain connected. 
The \emph{basis} occurs before any deletion and holds trivially. 
At an induction \emph{step}, consider dealing with player $i \in C$.
Since $i \in C$,
$\alpha_p \geq \alpha_i$, and since for each $j \in C$,
the induction hypothesis of the connectedness of $C$ implies
$\deg(j) \geq p + m - 1 \geq \alpha_p \geq \alpha_i$, $i$ does not 
want to disconnect from $j$. The second inequality stems from the definition of~$p$.

Second, we prove that all the players not in $C$ become isolated during the execution of Algorithm~\ref{alg:max_incled_ne}.
If not, let $i$ be the maximum $i$ that remains not isolated
after Algorithm~\ref{alg:max_incled_ne} terminates. Then,
from $i$'s maximality, $i$ sees at most $i - 1$ connected
players and $m$ connected non-players, which all are
completely interconnected, therefore have the degree of $i - 1 + m$
each. Since $i \not \in C \Rightarrow i + m - 1 < \alpha_i$,
and so $i$ prefers to disconnect, contradicting the definition
of Algorithm~\ref{alg:max_incled_ne}.
\end{proof}

We now consider the efficiency of the greatest \PANS.
If $p > 0$, then the utility of any player~$i \in \set{1, \ldots, p}$
is $(p + m - 1) (p + m - 1) - \alpha_i (p + m - 1)
= (p + m - 1 - \alpha_i) (p + m - 1)$, while and all
the other players obtain the utility zero.
Thus, the social welfare is 
$(p + m - 1) \sum_{i = 1}^p{(p + m - 1 - \alpha_i)}$.
On the other hand, if $p = 0$, then every player obtains 
the utility zero, resulting in $\sw = 0$.

Next, we characterise the smallest \PANE. We first find a
qualitative characterisation.
\begin{proposition}\label{prop:least_pane_char}
The resulting graph $G'$ returned by Algorithm~\ref{alg:least}
is either identical to the original graph $G_0$, or $G'$
consists of a fully connected set $C \supseteq V \setminus N$, 
while all the nodes $V \setminus C$ are isolated, that is
$E' = \set{(j, l) : j, l \in C}$.

\end{proposition}
The proof performs an induction on the algorithm sweeping over nodes
with various $\alpha$s.

\begin{proof}
Since the result of Algorithm~\ref{alg:least} is independent of the
order of visiting the nodes, we assume the nodes are visited
in the order $1, 2, \ldots, n$.
We prove by induction on a possible execution of Algorithm~\ref{alg:least}
that we always have either $G_0$, or a fully connected component including
$V \setminus N$ and all the nodes outside this component are isolated.
Denote this component after considering all the agents with
$\alpha_i < t$ by $C_t$, and if the graph is still the original one,
denote $C_t \defas \emptyset$.

At the \emph{induction basis}, consider all the players $\set{i \in N : \alpha_i < \max\set{1, m}}$.
Algorithm~\ref{alg:least} will interconnect
them together with all the $V \setminus N$ 
in a clique $C_{\max\set{1, m}}$.
Since there may be no such players, $C_{\max\set{1, m}}$ may be empty. 
If it is, the algorithm terminates, resulting in the original graph $G_0$.
Otherwise, the execution continues as described next.

At an \emph{induction step}, let $C_t \neq \emptyset$ be the currently attained
clique. Let $N_t \defas \set{i \in N \setminus C_t : \alpha_i < \abs{C_t}}$.
If $N_t \neq \emptyset$, then each one of $N_t$ will strictly benefit
from connecting to all the nodes in $C_t$. Then, each node in of $N_t$ will also strictly benefit from connecting to all the other nodes of $N_t$, as it will then possess the degree of at least $\abs{C_t} + 1$.
Then, each node in $C_t$ will strictly benefit from connecting to each node
of $N_t$, because the degree of each node in $N_t$ will then be 
even greater than the degree of $C_t$, which in turn was enough to
motivate connection.

Thereby, be maintain the invariant, extending the clique to be
$C_{\abs{C_t}} \defas C_t \cup N_t$. We continue in this manner,
till $N_t = \emptyset$, at which point no node or pair of nodes
want to create a connection. Indeed, a player~$i$ in $N \setminus C_t$
has $\alpha_i \geq \abs{C_t} \geq 1$, thus $i$ wants no connection.
\end{proof}

We will now find the size of clique $C$ from
Proposition~\ref{prop:least_pane_char}.
\begin{proposition}\label{prop:least_pane_char_size}
Let $q \defas \min\set{i \in N : \alpha_i \geq \max\set{1, i + m - 1}}$, where
$\min\emptyset \defas n + 1$.
The resulting graph $G' = (V, E')$ returned by Algorithm~\ref{alg:least}
fulfills the following. If $q = 1$, then $G' = G_0$, i.e., all the players are disconnected. Otherwise, there exists a fully connected component from nodes
$C \defas \set{1, \ldots, q - 1} \cup (V \setminus N)$,
and all the other nodes are independent, meaning that
$E' = \set{(j, l) : j, l \in C}$.
\end{proposition}
\begin{proof}
The inductive proof of Proposition~\ref{prop:least_pane_char}
stops at the first time that $\alpha_i$ is at least the size of
the clique made from players $1, 2, \ldots, i - 1$ (or just $1$) 
and all the non-players~$V \setminus N$. Thus, the player on
which the growth of the clique stops is $q$.
\end{proof}

By analogy to the efficiency of the greatest $\PANS$, the efficiency
of the smallest one is calculated as follows. If $q > 1$, then
the utility of any player $i \in \set{1, \ldots, q - 1}$ is
$(q + m - 2 - \alpha_i) (q + m - 2)$, and all the other players
receive the utility zero. Therefore, the social welfare is
$(q + m - 2) \sum_{i = 1}^{q - 1}{(q + m - 2 - \alpha_i)}$.
If $q = 1$, then everyone attains the utility zero, and thus
$\sw = 0$.

We will next connect efficiency with edge inclusion and
with the strength of PANE{}.
Building upon these results, we will find the prices of anarchy 
and stability and characterise the effective uniqueness of \PANE{}. 

%
%
\subsectioninline{Social welfare and edge inclusion}
\label{Sec:hiders_game:eff_incl}
We first relate the social welfare to the inclusion
of the resulting edge sets. 

Given any node $i$ in $V$, player or not,
let \defined{the player degree, $\deg'$,} of~$i$ be the number of players in
the node's neighbourhood, i.e.~$\deg'(i) \defas \abs{N(i) \cap N}$.

\begin{theorem}\label{the:mono_util_incl}
If Nash stable (\NS{}) graph $F = (V, E_F)$ strictly \mbox{(edge-)}includes $G = (V, E_G)$, meaning 
$E_F \supsetneq E_G$, then, for each player $i$,
$u_i(F) \geq u_i(G)$. 

Furthermore, the equality holds for each player
if and only if for each edge $e = (i, j) \in E_F \setminus E_G$, 
all the following conditions hold:
\begin{enumerate}\label{u_eq_cond}
	\item \label{u_eq_cond:G}
	In $G: \deg'_G(i) = \deg'_G(j) = 0$.
	
	\item	\label{u_eq_cond:F}
	In $F$:
	$j \in N \Rightarrow
	\deg_F(i) = \alpha_j$ and $i \in N \Rightarrow \deg_F(j) = \alpha_i$.
\end{enumerate}
\end{theorem}
In Example~\ref{example:subsemilatt_non_mono}, this implies that $G_3$
is the \emph{only} Nash stable graph with the maximum social welfare,
since all the $\alpha$s are non-integer, thus cannot be equal to 
any player's player degree.

The following proof uses the equilibrium property to prove the main inequality. The characterisation is then proven employing the equilibrium conditions and complementarity to edge addition.

\begin{proof}
Starting from $F$, fix any player $i$ and
let $i$ delete $\set{(i, j) : j \in \Neighb_F(i) \setminus \Neighb_G(i)}$, 
also
deleting the edges that have been added \emph{only} by $i$ between his neighbours, when at least one ceases being a neighbour, that is $i$ stops doing
$\set{\cnct_i(j, l) : j \in \Neighb_F(i) \setminus \Neighb_G(i) \text{ or } l \in \Neighb_F(i) \setminus \Neighb_G(i)}$, 
and no other player~$k$ does 
$\cnct_k(j, l)$, where $j \in \Neighb_F(k)$ and $l \in \Neighb_F(k)$.
Denote the resulting graph by $F'$.
Then, $u_i(F) \geq u_i(F')$ since $F$ is an \NE. Additionally,
$u_i(F') \geq u_i(G)$, since $\deg_{F'}(i) = \deg_{G}(i)$, and the degrees of its neighbours in $F'$ are at least
as large as they are in $G$, meaning that $\deg_{F'}(j) \geq \deg_{G}(j), \forall j \in \Neighb_G(i)$. Transitivity implies 
$u_i(F) \geq u_i(G)$.

We now approach the question: ``When is every player's utility 
equal in $F$ and $G$?'' The intuition is that condition~\ref{u_eq_cond:G} of the theorem
cares for the new neighbours of the old neighbours, while 
condition~\ref{u_eq_cond:F} relates to the new neighbours.
First, we assume that the conditions above hold and prove 
that $\forall i \in N, u_i(F) = u_i(G)$. 
Indeed, fix any
player~$k$, and add edges from $G$ to $F$, while adding the edges in $(E_f \setminus E_G) \cap \set{(k, y) : y \in \Neighb_F(k)}$ last. Then, $u_k$ is constant till the edges $(E_f \setminus E_G) \cap \set{(k, y) : y \in \Neighb_F(k)}$ start being added, since $k$ cannot neighbour in $G$ any
node that may be added an edge, by the assumption $\deg'_G(i) = \deg'_G(j) = 0, \forall e = (i, j) \in E_F \setminus E_G$. Now, once some edge $(k, i)$ is added, $u_k$ stays constant, since we assume $\deg_F(i) = \alpha_k$.

Conversely, assume that $u_i(F) = u_i(G), \forall i \in N$. 
We first derive the second condition.
Since $F \supsetneq G$, let $e = (i, j) \in F \setminus G$
and denote $\hat F \defas F \setminus \set{e}$ (assume no
edges have been added by a player whom $e$ made adjacent to a
non-player, since we can always add such edges to $E_0$ w.l.o.g.,
because all the assumptions will hold, too, and so it will suffice to prove the required there). 
Now, let $r$ be any player among $i$ and $j$, and let $r'$ be the other node
(in particular, both can be players). 
The equilibrium conditions imply $\deg_F(r') \geq \alpha_r$.
If this inequality was strong, then the complementarity to edge addition would imply $u_r(F) > u_r(G)$, contradictory to the assumption. We infer, $\deg_F(r') = \alpha_r$, the second condition.

Let us now prove the first condition.
Consider any $e = (i, j) \in E_F \setminus E_G$ and
first, assume by contradiction that $\deg_G'(i) > 0$, meaning that $\exists k \in N, (i, k) \in E_G$.
Then, create $F'$ from $F$ by deleting all the
edges $\set{(k, y) : y \in \Neighb_F(k)} \setminus \set{(k, y) : y \in \Neighb_G(k)}$, and all the edges between the non-player neighbours of
$k$, connected only by him, if at least one of which is not a neighbour any more, i.e.~$k$ stops doing $\set{\cnct_k(r, l) : r \in \Neighb_F(k) \setminus \Neighb_G(k) \text{ or } l \in \Neighb_F(k) \setminus \Neighb_G(k)}$. We then obtain
$u_k(F') > u_k(G)$, assuming that $(i, j) \in E_{F'}$. Additionally,
$F \in \NS \Rightarrow u_k(F) \geq u_k(F')$. Therefore, $u_k(F) \geq u_k(F') > u_k(G)$,
contradictory to the assumption of the same utilities in $F$ and $G$.
On the other hand, if $i$ becomes disconnected from $j$ in $F'$, 
that is if $(i, j) \in E_F \setminus E_{F'}$, then
we conclude that $(k, j) \in E_F \setminus E_{F'}$.
Now, since we have proven that $\deg_F(j) = \alpha_i$, player~$k$ has not gained from deleting $(k, j)$, even before it deleted $(i, j)$, so after deleting $(i, j)$, player~$k$ strictly loses, namely $u_k(F) > u_k(F')$. Since $u_k(F') \geq u_k(G)$, we conclude $u_k(F) > u_k(G)$, contradicting the assumption again.
%
\end{proof}

This theorem implies the (inclusion-wise) greatest and the least \PANS{} attain the greatest and the least social welfare, respectively. Together with Propositions~\ref{prop:great_pane_char} and
\ref{prop:least_pane_char} about the structure
of the greatest and the least \PANS{}, we can find the $\poa$ and the $\pos$, provided we know 
the optimal social welfare. 

\subsectioninline{Strength and efficiency}
\label{Sec:hiders_game:stren_eff}
The proven theorem allows characterising strength of \PANS{} by maximum
utilities.
\begin{theorem}\label{the:max_util_iff_strong}
For any $k \geq 1$, a pairwise $k$-strong Nash stable graph $F$ is 
of maximum utility among the $k$-strong \PANS{} for each player iff it is strong.%
\footnote{Pairwise Nash is $1$-strong and strong pairwise
Nash is $n$-strong.}
\end{theorem}
The proof first connects non-maximality of utility
with a profitable deviation, implying the theorem for $k = 1$.
This claim allows proving the theorem for any $k \geq 1$.

\begin{proof}
First, we prove the theorem for $k = 1$. Assume first that $G$ is strong. 
Algorithm~\ref{alg:great} returns a \PANS{}, say $F$, that is also
actually strong, by Proposition~\ref{prop:unil_del_enough}. The utility
monotonicity Theorem~\ref{the:mono_util_incl} implies that each player
attains her maximum utility in $F$, so unless each
player obtains her maximum utility in $G$, profile $F$ can be seen as
a profitable deviation of all the players from $G$, contradicting its
assumed strength.

Conversely, assume that in $G$, each player obtains her maximum utility.
Let $F$ be the profile returned by Algorithm~\ref{alg:great}.
Consider any set deviation from $G$ and the same deviation from $F$.
Consider any edge that is present in $F$ but not in $G$, say $e = (i, j)$.
Now, for any set deviation from $F$, since no player neighbours
$i$ or $j$ in $G$, by Theorem~\ref{the:mono_util_incl},
the only players that might benefit
less than they would benefit if the deviation was from $G$ could be
$\set{i, j} \cap N$, but they could just disconnect, if desired.
Therefore, since $F$ does not have a profitable deviation, neither
does $G$. Thus, $G$ is strong.

Having proven the theorem for $k = 1$, we consider a general $k > 1$.
We have shown that a strong \PANS{} implies a maximum utility for
each player for \PANS{}, and therefore, either more so for
$k$-strong \PANS, which are a subset of \PANS{},
by Theorem~\ref{The:strong_pane_not_empt_subset_incl}.
Conversely, a maximum utility $k$-strong \PANS{} is, by the proven above,
also a maximum utility among the \PANS{}, and this, as we have
demonstrated, implies strength.
\end{proof}

We readily conclude the following about strong \PANS.
\begin{corollary}\label{cor:same_util_if_strong}
For any $s, t \in n-\PANS, \forall i \in N, u_i(s) = u_i(t)$.
\end{corollary}

We are now all set to characterise the effective uniqueness of pairwise Nash equilibria,
i.e.~the uniqueness of utilities in
all the \PANE.
Interestingly, this is done through strength.
\begin{proposition}\label{prop:unique_util_char}
For any $k \geq 1$, the utility of each player is the same in all the $k$-strong \PANS{} if and only
if each k-strong \PANS{} is also strong. In other words, 
for any $k \geq 1$, 
$$k-\poa = k-\pos \iff k-\PANS = n-\PANS.$$
\end{proposition}
\begin{proof}
Fix any given $k \geq 1$.
Now, each k-strong \PANS{} being strong is, by virtue of
Theorem~\ref{the:max_util_iff_strong}, equivalent to each $k$-strong
\PANS{} providing the maximum utility to each player, which is
equivalent to providing the same utility in all the $k$-strong \PANS.
\end{proof}

Theorem~\ref{the:mono_util_incl} implies that the inclusion-wise least \NE{},
which existence we show in Theorem~\ref{The:ne_not_empt_latt_incl},
has the minimum utility, while the greatest Nash equilibrium, which
existence also stems from Theorem~\ref{The:ne_not_empt_latt_incl},
has the maximum utility. Having the maximum utility for a pairwise \NE{} is equivalent to
being strong, as Theorem~\ref{the:max_util_iff_strong} states.
A natural question is whether a \PANS{} including another \PANS{}
has to be stronger; the answer is not, as
Example~\ref{example:subsemilatt_non_mono} demonstrates.
It is important to notice that the connections between strength
and efficiency we have seen in Theorem~\ref{the:max_util_iff_strong},
Corollary~\ref{cor:same_util_if_strong} and
Proposition~\ref{prop:unique_util_char}
do not hold in general strategic games, as we show in the following example.
\begin{example}
Consider the game with players $N = \set{1, \ldots, n}$, player~$i$'s
strategies $S_i = \set{1, \ldots, n}$ and her utility defined as
$$u_i(s) \defas 
\begin{cases}
n - \abs{i - j} + \frac{j}{10 n}, & \textbf{If } \exists j \in N, \text{s.t. } s_1 = \ldots = s_n = j,  \\
0,  															& \text{Otherwise}. \\
\end{cases}$$

In this game, we make the following observations:
\begin{itemize}
	\item $\PANE = n-\PANE$ (all the \PANE{} are strong);
	\item	The utilities of the players and even the $\sw{}$ differ among
	the various \PANE;
	\item	$\poa = k-\poa < 1$, while $\pos = k-\pos = 1$.
\end{itemize}

\end{example}

\subsectioninline{Efficiency bounds}
\label{Sec:hiders_game:eff_bound}
Generally, the social welfare
constrained by the pairwise equilibrium requirement can differ infinitely
from the actual optimum, as the following example demonstrates.
\begin{example}\label{ex:inf_k_pos}
Consider a graph on two players, $1$ and $2$, such that
$\alpha_1 = 0, \alpha_2 = 1 + \epsilon$, for a positive $\epsilon$.
Here, the only (strong) \PANS{} is the empty graph, because player $2$
would not like to connect, and the social welfare of the empty graph is zero. The optimum
social welfare, however, is attained when the two players are connected,
and then the social welfare is $(1 - 0 \cdot 1) + (1 - (1 + \epsilon) \cdot 1)
= 1 - \epsilon$, which is positive for any $\epsilon < 1$.
Thus, even the price of stability is infinite here, so the predicted efficiency is very bad in this game. 
\end{example}

Thus, in order to obtain better bounds on the efficiency loss, we need
to introduce additional conditions, which we do in the next section.
In the general case, we bound the additive loss in social
welfare as follows.
\begin{proposition}\label{prop:gen_add_bound}
$\max_{s \in S}\sw(s) - \min_{t \in \NE}\sw(t) \leq$ 
$n \frac{(n - 1) (n - 2)}{2}
+ \abs{\set{i \in N : \alpha_i < n + m - 1}} (n - 1) (n + m - 1)
- \set{ \sum_{i : \alpha_i < n + m - 1}{\alpha_i}}
+ n m (n - 1 + n m)$.
\end{proposition}
The idea is first to prove that given any \NE{} graph $G$,
the optimum social welfare can be reached in a graph including $G$.
Then, we bound the utility difference per each edge. 

\section{Full Characterisation of 2 Classes}\label{Sec:hiders_cases}

Intuitively, Example~\ref{ex:inf_k_pos} yields
zero social welfare in the only pairwise stable equilibrium, because
the only connection was undesirable for one side, but the total utility
it provides is positive. This suggests limiting the space of 
coefficients $\alpha_i$ to bound the inefficiency of \PANS{}.

We now complete characterise the \PANS{} and find the prices of anarchy and stability for several salient classes of such games.
Analysing further important classes of games, such as when the number of non-players is much greater than the number of players, is also interesting.

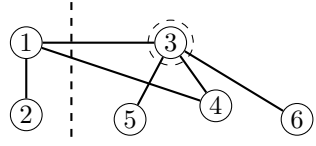
\begin{SCfigure}[1.2]
\begin{tikzpicture}[scale=0.6]
\tikzstyle{vertex} = [circle, draw=black, minimum size=12pt,inner sep=0pt]
\tikzstyle{title} = [circle, draw=white, minimum size=25pt,inner sep=0pt, font=\large\bfseries]
\tikzstyle{edge} = [-,thick]

\draw [dashed, thick, black] (0,1.3) -- (0,-1.8);
\draw [dashed, black] (2.7,0.4) arc [radius=0.5, start angle=0, end angle= 360];

\node[vertex](v1) at (-1, 0.4){1};
\node[vertex](v2) at (-1, -1.2){2};



\node[vertex](vn1) at (2.2, 0.4){3};
\node[vertex](vn2) at (3.2, -1.0){4};


\node[vertex](vnm) at (1.3, -1.3){5};
\node[vertex](vn6) at (5.0, -1.3){6};

\draw[edge] (v1) -- (v2);
\draw[edge] (v1) -- (vn1);
\draw[edge] (v1) -- (vn2);
\draw[edge] (vn1) -- (vn2);
\draw[edge] (vn1) -- (vnm);
\draw[edge] (vn1) -- (vn6);

\end{tikzpicture}
\caption{Players~$N = \set{1,2}$ and non-players~$V \setminus N = \{3,4,5,6\}$.
$D_3(0) = 3$.
}%
\label{fig:Di0_def}
\end{SCfigure}

\subsectioninline{Equal $\alpha$s}\label{Sec:hiders_cases:eq_alpha}
Indeed, when all the players have the same 
need for privacy, meaning
all $\alpha_i$s are the same,
then we 
fully characterise all the pairwise Nash equilibria.
\begin{theorem}\label{the:ne_char:eq_alpha_m_0}
For $D \subset V \setminus N$ and a node $i \in D$, denote by $D_i(0)$
the number of edges from $i$ to $V \setminus N$ crossing the boundaries
of $D$ (such edges all have to belong to $G_0$), i.e.~%
$D_i(0) \defas \abs{\set{j \in V \setminus (N \cup D) | (i, j) \in E}}$.
Denote the smallest such number over nodes in $D$ by $D(0)$, i.e.~%
$D(0) \defas \min\set{D_i(0) | i \in D}$;
if $D = \emptyset$, define $D(0) \defas 0$.
(If $E_0 = \emptyset$, then
$\forall i \in D, D_i(0) = 0 \Rightarrow D(0) = 0$.)

If $\alpha \defas \alpha_1 = \ldots = \alpha_n$, 
then $G(s)$ of $s \in S$ fulfils the following:
\begin{enumerate}
	\item If $\alpha < \max\set{1, m}$, then $G(s)$ is \PANS{} if and only if it is complete.
	Such a \PANS{} is also strong.
		
	\item	If $\max\set{1, m} \leq \alpha \leq n + m- 1$, then $G(s) \in \PANS{}$
	if and only if $G(s)$ fulfils all the following conditions\footnote{These conditions include the graph $G(s) = G_0$.}:
	\begin{itemize}
		\item At most one non-empty connected component
		$D \subset V \setminus N$ connected to some players. If exists, $D$ is
		completely interconnected and
		$D(0) + \abs{D} \geq \alpha + 1 - n$ (possible, since $\alpha + 1 - n \leq m$)
		and $(i, j) \in E, \forall i \in N, j \in D$.
		\item	The induced subgraph $G(s)[N]$ has at most one non-empty connected
		component $C$, which, if exists, is completely interconnected and
		$\abs{C} \geq \alpha + 1 - \abs{D}$ (this is possible, because
		$\alpha + 1 - \abs{D} \leq n$).
		\item	Additionally, if $\alpha < m + 1$ and $D = V \setminus N$,
		then $C = N$.
	\end{itemize}

	\item	If $n + m - 1 < \alpha$, then $G(s)$ is \PANS{} if and only if it is
	equal to $G$. 
	Such a \PANS{} is also strong.
\end{enumerate}
\end{theorem}
\figref{fig:Di0_def} illustrates the definition of $D_i(0)$ in the statement of this theorem.

The larger the visibility aversion
is, the less connected the resulting \PANS{} becomes.
Consider an example without non-players.

\begin{example}[$m  =0$]\label{ex:m_0}
Consider the example of $n$ players with the coefficient $\alpha$
and no other nodes, meaning that every node is strategic, i.e.~~$m = 0$.
In the notation of the theorem, each $D = \emptyset$, and thus,
$D(0) = 0$. Therefore, our theorem implies the following:
\begin{enumerate}
    \item If $\alpha < 1$, then the only \PANS{} is the complete graph,
    and it is also a strong \PANS.
    
    \item If $1 \leq \alpha \leq n - 1$, then the only \PANS{}
    has at most one non-empty connected component, which, if exists,
    if completely connected and contains at least $\alpha + 1$
    players.
    
    \item If $n - 1 \leq \alpha$, then a graph is \PANS{}
    if and only no connections take place. This \PANS{} is
    strong.
\end{enumerate}
\end{example}

We now prove the theorem.

\begin{proof}
Assume profile $s$ is a \PANE{} and consider the resulting graph
$G(s)$. 
If $i \in N$ is connected to $j \not \in N$, then 
$\deg(j) \geq \alpha$, so any player is connected to $j$; there can be
no player disconnected from $j$, since connecting that player would make
$j$'s degree strictly larger than $\alpha$, strictly benefiting that
player.
In addition, if $j, k \not \in N$ are connected to some players,
then since they are connected to all the  players each, as we have
demonstrated, we conclude that they will be connected as well in
any \PANS. Thus, all the non-players that are connected to some players, are
connected to \emph{all} the players each and \emph{completely} interconnected
among themselves.

%

We are now set to prove the statements. First, we assume we have a \PANS{}
and prove the stated conditions take place.
If $\alpha < m$, then each player would profit from connecting to all the
non-players and interconnecting them, since then each one of the
non-players will have the degree of at least $m$, which implies that 
for any profile $s \in S$ where $\exists i \in N$ such that not all
non-players are connected to $i$ or not all of them are interconnected,
the deviation $$s'_i \defas s_i \cup \bigcup_{j \in V \setminus N}{\cnct(i, j)}
\cup \bigcup_{i \neq k \in V \setminus N}{\cnct_i(j, k)}$$ would benefit $i$,
i.e.~$u_i(s'_i, s_{-i}) > u_i(s)$. Since this would make
each player's degree at least $m$, each pair of players would profit from
complete interconnecting. Thus, the graph has to be complete.
If we just have that $\alpha < 1$, we conclude that each
player would profit from connecting to another node, implying completeness.

On the other hand, let us assume
$\max\set{1, m} \leq \alpha \leq n + m - 1$.
If the graph is equal to $G$, all the conditions are fulfilled.
Assuming some edges have been added  to $G$, denote all the nodes
in $V \setminus N$ that are connected to some players by $D$ ($D$ could
be empty as well). To avoid the players having
profit from disconnecting, we need the degree of any node~$i$ in $D$ to be
at least $\alpha$, i.e.~$\forall i \in D: D_i(0) + \abs{D} - 1 + n \geq \alpha
\iff D(0) + \abs{D} \geq \alpha + 1 - n$. This holds for the empty  $D$ too,
since then, $0 \geq \alpha + 1 - n$, because $n - 1 \geq \alpha$, for
otherwise, no edges may have been added to $G$, because the non-players are not
connected to the players ($D = \emptyset$) and there exist only $n$ players.
Additionally, as shown above, $D$ is completely interconnected and
$(i, j) \in E, \forall i \in N, j \in D$.
Now, consider the induced graph $G(s)[N]$. Intuitively, we now discard
the uniform influence of the nodes in $D$ on the others.
The uniform influence of $D$ consists of being connected to its $\abs{D}$
interconnected nodes.
If any two players $i, j \in N$ are connected and so are players
$k$ and $l$, then the \NE{} condition implies that
$\deg_{G(s)[N]}(i) + \abs{D} \geq \alpha$ and $\deg_{G(s)[N]}(j) + \abs{D} \geq \alpha$ and, for
the other pair, we have
$\deg(k)_{G(s)[N]} + \abs{D} \geq \alpha$ and $\deg(l)_{G(s)[N]} + \abs{D} \geq \alpha$. Therefore, 
the assumption of a \PANE{} implies that $i$ is connected to $k$ and to $l$
and so is $j$, 
and $k$ is connected to $i$ and to $j$, and so is $l$;
the necessity strictness stems from the fact that if some of those connections did not exist, a new connection would increase the
degrees, implying a strict profit. 
Therefore, there exists at most one non-empty connected component
in $G(s)[N]$, which, if exists, we call $C$, and it is completely connected.
If exists, the component $C$ has to fulfil $\abs{C} - 1 + \abs{D} \geq \alpha
\iff \abs{C} \geq \alpha + 1 - \abs{D}$, to avoid the players having
profit from disconnecting.
%

In the particular case of $\alpha < m + 1$ and $D = V \setminus N$, 
since each player is connected to all the nodes in $D$, each player has
$m$ non-player neighbours, and since $\alpha < m + 1$, having a \PANS{}
implies all the players are interconnected, meaning $C = N$.

Finally, if $n + m - 1 < \alpha$, then disconnection is always strictly
profitable, so the original graph $G$ is the only \NS{} graph.

In the other direction, the conditions imply pairwise Nash stability
as follows.
For $\alpha < m$, we have showed that each player benefits from
the complete interconnection. This is a strong \PANS{} too, as
no player would strictly benefit in any other graph, by
Theorem~\ref{the:mono_util_incl}.

For $\max\set{1, m} \leq \alpha \leq n + m- 1$, each player faces
either nodes of $D$ with degree $D_i(0) + n + \abs{D} - 1 \geq \alpha$ or players
of $C$ with degree $\abs{D} + \abs{C} - 1 \geq \alpha$, and therefore he
does not want to disconnect. No player
would profit from connecting to a new node as well, because any
non-player $j \in V \setminus (N \cup D)$ would receive the degree
$1 + \set{j}_j(0) \leq m \leq \max\set{1, m} \leq \alpha$, the first
inequality holding unless $\set{j}_j(0) > m - 1$, which contradicts
$\set{j}_j(0)$ being the number of neighbours of $j \in D$ in
$V \setminus (N \cup D)$.
Now, the not connected players
would obtain the degree of $\abs{D} + 1 \leq m + 1$, which is bounded by
$\alpha$, unless $D = V \setminus N$ and $\alpha < m + 1$, in which case all the players are
connected.
Thus, this is a \PANS.

Finally, for $n + m -1 < \alpha$, no coalition would like to connect,
since any nodes' degrees will become at most $n + m -1$ after the connection, which is
less than $\alpha$.
\end{proof}

Theorem~\ref{the:ne_char:eq_alpha_m_0} allows us prove the
following clean efficiency result.
\begin{theorem}\label{the:ne_char:eq_alpha_m_0:eff}
If $\alpha_1 = \alpha_2 = \ldots = \alpha_n$ (let $\alpha \defas \alpha_i$),
then the following holds for the resulting graphs:
$$
\begin{cases}
k-\poa = k-\pos = 1, \forall k \geq 1	&	\textbf{if } \alpha < \max\set{1, m};\\
\pos = 1, \poa = \infty &	\textbf{if } \max\set{1, m} \leq \alpha \leq n + m - 1;\\
k-\poa = k-\pos = 1, \forall k \geq 1 & \textbf{if } n + m - 1 < \alpha.\\
\end{cases}
$$
\end{theorem}
Therefore, if $\max\set{1, m} \leq \alpha \leq n + m - 1$,
both very efficiency and very inefficient situations can occur, and the players should regulate how they act.
\begin{example}
In Example~\ref{ex:m_0}, this theorem implies the following:
$$
\begin{cases}
k-\poa = k-\pos = 1, \forall k \geq 1	&	\textbf{if } \alpha < 1;\\
\pos = 1, \poa = \infty &	\textbf{if } 1 \leq \alpha \leq n - 1;\\
k-\poa = k-\pos = 1, \forall k \geq 1 & \textbf{if } n - 1 < \alpha.\\
\end{cases}
$$
\end{example}

The proof of the theorem follows.

\begin{proof}
First, we compute the maximum social welfare in a social welfare
maximising resulting graph $G(s)$.
In general,
the social welfare is the total utility, which is ($\deg'(i)$ denotes
the number of neighbours in $N$, i.e. $\abs{\Neighb(i) \cap N}$)
\begin{eqnarray}
\sum_{i \in N}{u_i(A, D)} = \sum_{i \in N}{\paren{ \sum_{j \in \Neighb(i)}{\deg(j)} - \alpha_i \deg(i) }}\nonumber\\
= \sum_{i \in V}{\paren{\deg'(i) \deg(i)}} -  \sum_{i \in N}{\paren{\alpha_i \deg(i) }}.
\label{eq:sw_form}
\end{eqnarray}

If $\alpha > n + m - 1$, then the maximum social welfare is achieved
by adding no connections. Otherwise, we can assume w.l.o.g.\ all the
players are connected to all the non-players, and all the non-players
are completely interconnected among themselves. Thus, for
$\alpha \leq n + m - 1$,
for each $i \in N$ holds $\deg(i) = \deg'(i) + m$, since each
player is connected to all the $m$ nodes in $V \setminus N$, and for each
$j \in V \setminus N$, we also have $\deg'(j) = n$.
Therefore, \eqnsref{eq:sw_form} becomes
\begin{eqnarray*}
\sum_{i \in V}{\paren{\deg'(i)^2 + m \deg'(i)}} - \alpha \sum_{i \in N}{\paren{\deg'(i) }} - \alpha n m\\
= \sum_{i \in N}{\paren{\deg'(i)^2 + m \deg'(i)}} + m n (n + m) - \alpha \sum_{i \in N}{\paren{\deg'(i) }} - \alpha n m.
\end{eqnarray*}
For each $i \in N$, we analyse the $i$th term of this expression for maximum,
assuming $\deg'(i)$ can be chosen freely, regardless the other degrees
and not even necessarily integer. We will see that even the
optimum under these assumptions can be attained. 
\begin{eqnarray*}
\frac{\partial(\deg'(i)^2 + m \deg'(i) -  \alpha \deg'(i) - \alpha m)}{\partial(deg'(i))} 
= {2 \deg'(i)+ m -  \alpha}
\Rightarrow \\ \frac{\partial(\deg'(i)^2 + m \deg'(i) -  \alpha \deg'(i) - \alpha m)}{\partial(deg'(i))} = 0
\iff \deg'(i) = \frac{\alpha - m}{2}.
\end{eqnarray*}
Since the second derivative is $2$, which is (strictly) positive,
$\frac{\alpha - m}{2}$ is a local minimum. Since this is a convex parabola,
$\frac{\alpha - m}{2}$ is a global minimum as well, and the global maxima are
attained at the extreme values of the domain. This holds if
$\frac{\alpha - m}{2}$ belongs to the domain, i.e.~$[0, n - 1]$.
Otherwise, the global maxima also are attained at
the extreme values of the domain. For $\deg'(i) = 0, \forall i \in N$,
the social welfare is $m n (n + m) - \alpha n m$, while 
for $\deg'(i) = n - 1, \forall i \in N$, which is the maximum possible
$\deg'(i)$, the social welfare is
$n ((n - 1)^2 + m (n - 1)) + m n (n + m) - \alpha n (n - 1) - \alpha n m
= n (n - 1) (n + m - 1 - \alpha) + m n (n + m) - \alpha n m$, 
which dominates $m n (n + m) - \alpha n m$ for $\alpha \leq n + m - 1$.
To conclude, the maximum social welfare is 
\begin{eqnarray}
\begin{cases}
    n (n - 1) (n + m - 1 - \alpha)  & \textbf{for } \alpha \leq n + m - 1; \\
    + m n (n + m) - \alpha n m & \\
    0 & \textbf{for } n + m - 1 < \alpha,
  \end{cases}
\label{eq:sw_max}
\end{eqnarray}
zero stems from adding no connections, as we mentioned above.

We now prove the statement case by case.
For $\alpha \leq n + m - 1$, \eqnsref{eq:sw_max} implies that the
complete graph maximises the social welfare. 
Now, if $\alpha < \max\set{1, m}$, then
Theorem~\ref{the:ne_char:eq_alpha_m_0} states that the only \PANS{} is
the complete graph. 
Therefore, $\poa = \pos = 1$.
Since the complete graph is also a strong \PANS, $k-\poa = k-\pos = 1$,
for each $k \geq 1$.

On the other hand, if $\max\set{1, m} \leq \alpha \leq n + m- 1$, then
Theorem~\ref{the:ne_char:eq_alpha_m_0} allows for the complete graph as
a \PANS, which implies $\pos = 1$.
However, we can have other \PANS{} as well, such as adding no connections,
and so $\poa = \infty$.

Finally, if $n + m - 1 < \alpha$, then \eqnsref{eq:sw_max} states that
zero is the maximum possible social welfare.
Theorem~\ref{the:ne_char:eq_alpha_m_0}, in turn, states that s graph is
a \PANS{} if and only if no edges are added, and this \PANS{} is also
strong.
The social welfare of this profile is zero, and
therefore, $k-\poa = k-\pos = 1$,
for each $k \geq 1$.
\end{proof}

Using Theorem~\ref{the:ne_char:eq_alpha_m_0:eff} and
Proposition~\ref{prop:unique_util_char}, we conclude the following about
the strength of the \PANS{} in Theorem~\ref{the:ne_char:eq_alpha_m_0},
when $\max\set{1, m} \leq \alpha \leq n + m- 1$.
\begin{corollary}\label{cor:ne_char:eq_alpha_m_0:str}
Under the conditions of Theorem~\ref{the:ne_char:eq_alpha_m_0}, if
$\max\set{1, m} \leq \alpha \leq n + m- 1$, there exist non-strong
\PANS.
\end{corollary}

%% file: hiders_cases.tex
\subsectioninline{Small $\alpha$s besides one}
\label{Sec:hiders_cases:sm_alpha_bes_1}
Consider the case when there exists one player, w.l.o.g.~player $1$, who cares about not
being too conspicuous differently, generally speaking, from the others, 
while the others do not care much,
i.e.~%
$\alpha_2, \alpha_3, \ldots, \alpha_n$ all are at most $\alpha$,
for some $\alpha < 1$.
Here, we provide the following characterisation.
\begin{theorem}\label{prop:pane_char:a1_others_eq}
Assume $\exists \alpha <1$, such that $\alpha_i \leq \alpha, \forall i = 2, \ldots, n$.
A graph is a \PANS{} if and only if all the following holds:
\begin{enumerate}
	\item all the nodes besides player $1$
	are completely interconnected,
	\item	and player $1$ is connected to 
	\begin{itemize}
		\item[if $\alpha_1 < n + m - 1 \Rightarrow$]	all the other nodes;
		\item[if $\alpha_1 = n + m - 1 \Rightarrow$]	to all the
		other players, while her connections to the non-players are arbitrary;
		\item[if $\alpha_1 > n + m - 1 \Rightarrow$]	no one.
	\end{itemize}
\end{enumerate}
The following change in the conditions yields the
characterisation of any $k$-strong \PANS{} for $k \geq 2$:
if $\alpha_1 = n + m - 1$, player $1$ has to be connected to all the
other nodes, including the non-players.
\end{theorem}
Thus, the more player~$1$ avoids visibility,
the less connected the resulting \PANS{} is. 
We reform
Example~\ref{ex:m_0} for these conditions. 
\begin{example}[$m = 0$]
Similarly to Example~\ref{ex:m_0}, consider the case of only
strategic nodes, meaning that $m = 0$. Then, the conditions of
the proposition on $\alpha_1 $ become
$\alpha_1 < n - 1, \alpha_1 = n - 1$ and $\alpha_1 > n - 1$.
\end{example}

And we now prove the stated theorem.

\begin{proof}
Consider any $s \in \PANE{}$ providing graph $G(s)$ from the original
graph $G$. First, all the players besides $1$ will connect to one another
and to all the non-player nodes, since $\alpha$ is strictly smaller than
$1$. They will also make all the non-player nodes connect to one another.
This means each node besides perhaps players~$1$ has the degree of at
least $n + m - 2$.
As for player $1$, he connects to all the other nodes if
$\alpha_1 < n + m - 2 + 1$ and does not connect if $\alpha_1 > n + m - 2 + 1$.
If the equality holds, he connects to all the other players, since
the player on the other side strictly benefits from such a connection.
As for connecting to the non-players, all the option are possible for $1$.

In the other direction, such a profile is an \PANS, since the players
others than $1$ definitely benefit from any connection, and player
$1$ benefits if $\alpha_1 < n + m - 1$, does not care
if the equality holds, and loses if $\alpha_1 > n + m - 1$.

Assuming $\alpha_1 = n + m - 1$ and
given any $k \geq 2$, in any $k$-strong \PANS, a coalition
of player~$1$ with another player~$i$ would not lose from $1$ connecting to all
the nodes, and $i$ would strictly profit. Therefore, the graph
has to be complete.
Conversely, assuming the conditions above, plus requiring that for
$\alpha_1 = n + m - 1$, all the nodes are interconnected,
such a \PANS{} has to be strong, for the following reasons.
First, no coalition $E \subseteq \set {2, 3, \ldots, n}$ can deviate
without (strictly) losing. When we also include in the coalition player $1$ who was connected
with the other nodes, he can only lose from disconnecting, regardless what
the other players do, unless $\alpha_1 = n + m - 1$, in which case $1$'s utility
may decrease or not change, while all the other players would strictly lose.
Finally, if $1$ was not connected, he would only lose from connecting,
regardless what the others do.
\end{proof}

Next, we analyse the efficiency of the characterised pairwise stable
graphs, assuming $\alpha_i$s for $i = 2, 3, \ldots, n$ are equal to some $\alpha$, in order to be able to provide exact formulas.
\begin{theorem}\label{prop:pane_char:a1_others_eq:eff}
Assuming that the $\alpha_i$s are:
$\alpha_1, \alpha_2 = \alpha_3 = \ldots = \alpha_n = \alpha$,
for some $\alpha < 1$, the prices of anarchy and stability are equal to 
(maximum social welfare)
$(n - 1)((n + m - 2)^2 - \alpha (n + m - 2))$ plus 
\begin{eqnarray*}
\begin{cases}
\max\set{0, (n - 1) (n + m -1 - \alpha) + n + m - 1 - \alpha_1}	&	\textbf{if } n + m - 1 > \alpha;\\
\max\set{0, n + m - 1 - \alpha_1} & \textbf{if } n + m - 1 \leq \alpha.\\
\end{cases}
\end{eqnarray*}
all this divided by (equilibrium social welfare)
\begin{itemize}
    \item[if $\alpha_1 < n + m - 1 \Rightarrow$]	$(n - 1)\set{(m + n - 1)^2 - \alpha (m + n - 1)} + (m + n - 1)^2 - \alpha_1 (m + n - 1)$;
    \item[if $\alpha_1 = n + m - 1 \Rightarrow$]	
        for $\poa$: $(n - 1)\set{(m + n - 1)^2 - \alpha (m + n - 1)} + (m + n - 1)^2 - \alpha_1 (m + n - 1)$,
        
        for $\pos$: $(n - 2)\set{(n - 2)(m + n - 1) + (n - 1) + m (m + n - 2) - \alpha (n + m - 1)} + (n - 1)(m + n - 1) - \alpha_1 (n - 1)$,
        
        only the bigger value remains for $k$-strong \PANS, for any $k \geq 2$.
    \item[if $\alpha_1 > n + m - 1 \Rightarrow$]	$(n - 1)\set{(m + n - 2)^2 - \alpha (m + n - 2)}$.
\end{itemize}
In particular, $\poa = \pos$ for $\alpha_1 < n + m - 1$,
for $\alpha_1 > n + m - 1$, or for $k$-strong \PANS, for any $k \geq 2$.
\end{theorem}

\begin{proof}
We first find the maximum social welfare. In order to maximise it,
players $2, \ldots, n$
have to be completely interconnected among one another and with all the
non-players, interconnecting them as well,
since each player can only benefit from
that. These interconnections already yield the social welfare of
$(n - 1)((n + m - 2)^2 - \alpha (n + m - 2))$.

As for player $1$'s connections, connecting it to $n' \leq n - 1$ other
players and $m' \leq m$ non-players yields each $i \in N$, such that
$(1, i) \in E(s)$, the extra utility of $n' + m' + (n' + m' - 1) - \alpha$.
Each $j \in N$ such that $(1, j) \not \in E(s)$ receives the added
utility of $n' + m'$, while player
$1$ obtains the extra utility of $(n' + m')(n + m - 1) - (n' + m') \alpha_1$.
Therefore, the total addition to the social welfare is the following
function of $n'$ and $m'$: $f(n', m') =
n' (n' + m' + (n' + m' - 1) - \alpha) + (n - n' - 1) (n' + m')
+ (n' + m')(n + m - 1) - (n' + m') \alpha_1
= n' (2n' + 2m' - \alpha - 1)
+ (n' + m') (2n - n' + m - 2 - \alpha_1)$.
obtained expression with respect to $n'$, as if it were a real parameter
in $[0, n - 1]$. 
\begin{eqnarray*}
\frac{d(n' (n' + m - \alpha) + n + m - 1 - \alpha_1)}{dn'}
= 2n' + m - \alpha.\\
\Rightarrow \frac{d(n' (n' + m - \alpha) + n + m - 1 - \alpha_1)}{dn'} = 0
\iff n' = \frac{\alpha - m}{2}.
\end{eqnarray*}
The second derivative is $2 > 0$, and so this is a single local and a
global minimum, and the global maxima are attained at $0$ or at $n - 1$.
At $n' = 0$, we have $n + m - 1 - \alpha_1$, while at $n' = n - 1$, we
obtain $(n - 1) (n + m -1 - \alpha) + n + m - 1 - \alpha_1$.
Thus, the maximum social welfare is 
$(n - 1)((n + m - 2)^2 - \alpha (n + m - 2))$ plus 
\begin{eqnarray*}
\begin{cases}
\max\set{0, (n - 1) (n + m -1 - \alpha) + n + m - 1 - \alpha_1}	&	\textbf{if } n + m - 1 > \alpha;\\
\max\set{0, n + m - 1 - \alpha_1} & \textbf{if } n + m - 1 \leq \alpha.\\
\end{cases}
\end{eqnarray*}

Having these formulas for the maximum social welfare, we now use the
characterisation of the \PANS{} from
Theorem~\ref{prop:pane_char:a1_others_eq} to calculate the social welfare in equilibria. So the social welfare in a \PANS{} is
\begin{itemize}
    \item[if $\alpha_1 < n + m - 1 \Rightarrow$]	$(n - 1)\set{(m + n - 1)^2 - \alpha (m + n - 1)} + (m + n - 1)^2 - \alpha_1 (m + n - 1)$;
    \item[if $\alpha_1 = n + m - 1 \Rightarrow$]	
        at most $(n - 1)\set{(m + n - 1)^2 - \alpha (m + n - 1)} + (m + n - 1)^2 - \alpha_1 (m + n - 1)$,
        
        and at least $(n - 2)\set{(n - 2)(m + n - 1) + (n - 1) + m (m + n - 2) - \alpha (n + m - 1)} + (n - 1)(m + n - 1) - \alpha_1 (n - 1)$,
        
        only the bigger value remains for $k$-strong \PANS, for any $k \geq 2$.
    \item[if $\alpha_1 > n + m - 1 \Rightarrow$]	$(n - 1)\set{(m + n - 2)^2 - \alpha (m + n - 2)}$.
\end{itemize}

And together with the formulas for the maximum social welfare, we obtain
the prices of anarchy and stability, and $k-\poa, k-\pos$, for any
$k \geq 2$.
\end{proof}

In the case of some players infiltrating a huge network, the model
predicts that the complete network will came to being. This result
shows that when dealing with certain cases, the model needs a extension
to express the bound on the number of possible connections.
\begin{proposition}
For large enough $m$ (i.e.~$m > \max\set{\alpha_i | i \in N}$),
the only \PANS{} is the complete graph. This is also a strong \PANS.

Here, $k-\poa = k-\pos = 1$.
\end{proposition}
\begin{proof}
If $m > \max\set{\alpha_i | i \in N}$, then any player would benefit
from connecting to all the non-players and completely interconnecting  
them, since that would make each non-player's degree at be least
$m > \alpha_i, \forall i \in N$. Therefore,
for any strategy profile $s$ where $\exists i \in N$ such that not all
non-players are connected to $i$ or not all of them are interconnected,
the deviation $s'_i \defas s_i \cup \cup_{j \in V \setminus N}{\cnct(i, j)}
\cup \cup_{i \neq k \in V \setminus N}{\cnct_i(j, k)}$ would benefit $i$,
i.e.~$u_i(s'_i, s_{-i}) > u_i(s)$. 
This, in turn, implies that any player's degree will be at least
$m > \alpha_i, \forall i \in N$, and so if in profile $t \in S$,
$\exists i \neq j \in N$, such that $(i, j) \not \in E(t)$, then
this is not pairwise stable, since
$u_i(t + (i, j)) > u_i(t), u_j(t + (i, j)) > u_j(t)$. Thus, a \PANS{}
has to be the complete graph.

Conversely, a complete graph is a strong \PANS, since no player will
benefit in any other graph. This also implies that the attained social
welfare is maximum, and therefore, $k-\poa = k-\pos = 1$.
\end{proof}

%% file: related_work.tex
\section{Related Work}\label{Sec:related_work}

Our work is related to a few bodies of research. We discuss them in relation to the key elements of our model.

\subsectioninline{Non-cooperative strategic models of network formation}
Perhaps the first \emph{non-cooperative} attempt to define stability was called \defined{Nash stable networks}%
~\cite{Myerson1991}, which is a Nash equilibrium in the
\defined{linking game}~\cite{Myerson1991}, where
each player declares her desired neighbours, and 
a link is formed
if and only if both players want it.
%

Since Nash stable networks can sustain too many cooperation
structures as \NE, such as the
empty graph with no connections, even when all the agents
want to interconnect, a number of refinements were proposed in the literature. In particular, Dutta et al.~\cite{DuttavandenNouwelandTijs1998}
analysed the equilibria of the game, where the
utilities are the
\emph{Myerson values~\cite{Myerson1977}.}
%
Similarly to
Myerson's linking game~\cite{Myerson1991}, they modelled
link formation as a strategic game, where 
a link is formed if and only if
both its sides want to connect. In order to avoid too many equilibria,
they turn to refinements of \NE,
namely undominated Nash equilibrium~\cite{DuttavandenNouwelandTijs1998}, coalition-proof Nash equilibrium,
and strong Nash equilibrium. 
The multiplicity of Nash equilibria here also
motivated refinements such as 
trembling hand and
proper equilibrium~\cite{CalvoArmengolIlkilic2009}. 
They provided many refinements, but provided no distinguished refinement.

Furthermore, having too many
Nash equilibria in the linking game%
~\cite{Myerson1991}
motivated
the classical pairwise stability notion by
Jackson and Wolinsky~\cite{JacksonWolinsky1996},
who consider a value function on the
graphs 
and an allocation rule.
They concentrate on two
cases of the model, namely the
\defined{connections model} and the \defined{co-author model},
the former inspiring our utility function by modelling gain from
being connected, and cost from direct connections. 
They define \defined{pairwise stability} of a graph,
requiring that any couple is connected only if
both players want to be connected, and is not connected only
if at least one player does not want to connect. This definition
sees the profile when two players who both want to connect
remain disconnected as not stable.
Jackson and Wolinsky 
show that 
sometimes no efficient graph is pairwise stable.
Unlike our paper,
they 
\emph{do not consider a non-cooperative game at all,
and they do not allow having non-active nodes or connecting other nodes.}
Belleflamme and Bloch%
~\cite{BelleflammeBloch2004} strengthen pairwise stability by allowing
to form or delete any number of
links, not only one link defined by a pair of nodes. 
Thereby, they add the notion of \NE{} to pairwise stability.
Among many follow-ups of Jackson and Wolinsky,
the truncated model~\cite{JacksonRogers2005} assumes
a partition of nodes, where
the costs are higher when the edge is between distinct equivalence class,
and no indirect benefits are obtained beyond a given distance.
In this model, pairwise stability and/or efficiency imply small world properties, which are short paths and high clustering. We employ the combined notion of \defined{Pairwise Nash stability}~\cite{BlochJackson2006},
requiring both pairwise stability and Nash stability. It disallows the
empty graph, when at least one connection is profitable to both sides.
This strengthening of pairwise stability is also used in
Belleflamme and Bloch%
~\cite{BelleflammeBloch2004}, 
who modelled the stability of collusive 
markets, and compared them to socially optimal 
networks. 
Additionally, pairwise Nash with transferable utilities has been considered in
~\cite{JacksonWolinsky1996,BlochJackson2006}. 
Goyal and Joshi~\cite{GoyalJoshi2003} studied pairwise Nash with
transferable utilities and they found asymmetric structures, like
stars, stable.
An illuminating comparison of the stability notions appears in~%
\cite{BlochJackson2006}.
Dutta and Mutuswami~\cite{DuttaMutuswami1997} defined a certain notion
of strong \NE{}, implying pairwise stability, and a weaker notion
of coalition proof \NE, which does not imply pairwise stability. They
subsequently connect allocation rules, stability and efficiency.
Following them, Jackson and van den Nouweland~\cite{JacksonvandenNouweland2005}
consider the strong \NE{} of Myerson's game~\cite{Myerson1991}, which is
more demanding than that of~\cite{DuttaMutuswami1997}, and
prove that its existence is equivalent to the non-emptiness of the core
in the derived cooperative game. They prove that if the strong \NE{}
of the Myerson's game (and equivalently the core) do exist, then that
strong Nash equilibria coincide with the efficient networks. 

\subsectioninline{Unilateral edge creation}
Our model and many others consider bilateral edge creation. Unlike
those, in the case of unilateral edge formation~\cite{BalaGoyal2000,Fabrikant2003,HojmanSzeidl2008}, Nash
equilibria is the natural stability concept, disallowing that two
players who both desire to connect remain disconnected.
Bala and Goyal~\cite{BalaGoyal2000} define
\defined{Directed Nash stability} as an \NE{} in a game where a node
rules the outgoing edges unilaterally; they
prove efficiency of many stable networks. 
They also
prove that the set of all the equilibria of this game is the absorbing
set of the learning dynamics, where a randomly chosen agent establishes
the best possible desired links at the moment.

\subsectioninline{Extensive games}
Different from the above one-shot models,
Aumann and Myerson~\cite{AumannMyerson1988} assume value division
using Myerson value~\cite{Myerson1977} and define 
a perfect information extensive game where nodes connect if
both want to. Compared to our model, this model lacks an action of connecting two other
nodes, does not allow for non-player nodes, and it features another utility
function. 
The advantage of such a model is that
the empty Nash equilibrium when no connections exist between players
who would benefit from interconnecting is not an \SPE.
The downside of extensive models includes the necessity to model
the intricacies of the interaction, diminishing the applicability
of such models.

\subsectioninline{Utilities}
Various utilities have been considered, such as the value from the
connected nodes minus the cost of the personally constructed
edges
~\cite{JacksonWolinsky1996,GoyalJoshi2006}, which has inspired us.
Bala and Goyal~\cite{BalaGoyal2000} define utility as the
connected component size minus the connection cost, while~%
\cite{GoyalJabbariKearnsKhannaMorgenstern2015} consider
reachability in the face of attack minus the connection and
immunity costs, connected components modelling connectivity
demands over the internet or in the society, while immunisation
modelling anti-virus software or vaccination. Our utility
complements that, the connection ability modelling
infiltration and underground organisations.
K\"onig et al.~\cite{KonigBattistonNapoletanoSchweitzer2011} set the
utility to be equal to the largest eigenvalue of the adjacency matrix
component of the agent minus the total cost of the adjacent edges.
Using the same utility definition, K\"onig et al.~%
\cite{KonigBattistonNapoletanoSchweitzer2012}
describe the efficient graphs by the means of the so called
\defined{nested stars} and then describe the stable graphs as well, so
that a comparison between efficiency and stability becomes possible.
The conclusion is that small and efficient networks are often stable,
while large, efficient ones are not.

\subsectioninline{Other models}
Unlike the complete knowledge models, such as ours,
the literature also studies information models 
such as different network knowledge~\cite{Gallo2012}.
There exist also various
efficiency models based on the possible allocations of the value to the
players, like
\defined{Pareto efficiency}~\cite[\sectn{5}]{Jackson2003}
or \defined{constraint efficiency}~\cite{Jackson2003b}, which 
are beyond the scope of this paper.


%% file: advice.tex
\section{Summary \& Policy Recommendations}\label{Sec:polic_rec}

We have shown that interaction of
players who aim to be hubs through their
neighbours, while keeping their own degree low,
is highly amenable to
analytical analysis. 
Aiming to avoid unrealistic equilibria, we
have considered pairwise Nash equilibria, including
the $k$-strong ones. 
These equilibria 
form a non-empty complete lattice, and the lattice of the
stronger pairwise equilibria is edge-included
in the lattice of the weaker \PANE.
Theorem~\ref{the:mono_util_incl} states that edge-inclusion
implies higher utility for all the players. 
The greatest possible graph exists and
is returned by Algorithm~\ref{alg:great}; 
this greatest \PANE{}
is strong as well as most profitable for all the players.
In general,
strength of pairwise equilibrium is equivalent
to maximum utility. 
These results suggest the following optimum policy for hiders and a
diagnostic tool for seekers.

\subsectioninline{A distributed algorithm for the players}
\label{Sec:advice:dist_player}
Theorem~\ref{the:mono_util_incl} implies that the both the total utility,
meaning the social welfare, and each player's utility are optimised when
the resulting graph is the greatest possible. Such a graph exists and
is returned by applying Algorithm~\ref{alg:great}. As an aside,
Proposition~\ref{prop:unil_del_enough} implies that this greatest \PANE{}
is strong as well. As for attaining this pairwise stable equilibrium,
instead of directly executing Algorithm~\ref{alg:great}, 
the characterising Proposition~\ref{prop:great_pane_char} suggests
each player calculate the maximum $i$ such that
$i + m - 1 \geq \alpha_i$. If no such
positive $i$ exists, then the optimum profile consists of the empty strategies,
while otherwise, denote this maximum $i$ by $p$ and return the clique
on nodes $\set{1, \ldots, p} \cup (V \setminus N)$, while leaving
nodes $\set{p + 1, \ldots, n}$ isolated.

If the $\alpha$'s are all equal or strictly
smaller than $1$, besides perhaps $1$ player, 
then the found
prices of anarchy and stability imply that
all the pairwise equilibria are socially optimal,
unless the agents fear visibility to a certain special extent,
in which case the pairwise equilibria can be either optimal
or infinitely inefficient.
This unnecessary inefficiency suggests
hiding agents regulate their behaviour.

All the above calculations and any regulation require
knowledge of the others' $\alpha$s, limiting the applicability.

\subsectioninline{Advising the external observer}
\label{Sec:advice:ext_obs}
Let us ask two questions:
\begin{enumerate}
    \item   Is the given graph infiltrated by hiding players?
    \item   Assuming infiltration, who are the players?
\end{enumerate}

Assuming the players follow the above strategies to 
attain the largest \PANE, Proposition~\ref{prop:great_pane_char}
implies that such an infiltrated graph consists of a clique and
isolated nodes. Assuming a realistic distribution of
the original graphs and on the parameters of the hiding players,
this graph is highly unlikely to occur ``naturally''. For example, if the distribution is scale-free~\cite{BarabasiAlbert1999}, then
the probability of a clique on $x$ nodes is approximately $1 / (x^\beta)$,
for a $\beta \in (2, 3)$.

Therefore, despite that
this method can only indicate the lack of infiltration when
the graph is not a clique plus isolated nodes, 
such a graph does indicate
infiltration with a high probability. Moreover, in such the case, we know that the not-connected
nodes are players, while nothing is known about the clique nodes. 
Of course, real agents are imperfectly rational, so we expect
graphs close to the above structure, rather than complete cliques and
fully isolated notes.

When all the agents care to avoid
visibility equally (meaning have equal $\alpha$'s), or when all the agents
besides one care perhaps not equally, but little, we 
fully characterise all the possible pairwise equilibria, not only the optimum ones. This,
in turn, allows to detect infiltration and locate the infiltrators even better.

The seeker should try to prevent potential perpetrators from forming
large interconnected networks, since they are most efficient.
For small networks, our results allow a seeker to calculate which pairwise
equilibria can form for any coalition size
(strength). 
Additionally, the provided bounds on how much social welfare
is lost in a pairwise equilibrium help estimate
the efficiency of the situation.

%% file: conclusion.tex
\section{Discussion \& Future Work}\label{Sec:conclusion}

As is, our model is limited to not too many non-players, since
given any set of players, a large enough number of
non-players will render the only pairwise equilibrium be
completely connected. A bound on the
number of connections or the submodularity of the gain from
being connected would solve this issue.
We also assume a complete information game, which makes sense in small
organisations. We could model incomplete information, departing from the relevant literature and requiring estimation of the knowledge players have.
Next, since a player may be unsure
whether the new acquaintance will strike root,
we can augment the model with probabilistic acquainting
neighbouring non-player nodes. We can also recall that real
non-player (i.e., not hiding) agents may still exhibit nontrivial behaviour, such as disconnecting or connecting. Modelling this would
render the model more cumbersome, but potentially more realistic.
Furthermore, in our model, 
we use the degree centrality as a proxy of the willingness of an agent to disclose information. Thus, implicitly, we assume uniform passing of information that does not have to be the case. To address this limitation, we could explicitly model strategic information transfer. This would increase the realism of the model for certain settings, but would render the model parameter-sensitive and less general. 
One can also try other centrality and visibility concepts and simulate the dynamics
of strategic interconnection, starting from the original graph, which is generated
according to some practical distribution.

An interesting alteration to our model would be considering the total number of the distinct neighbours of a node's neighbours, rather than the sum of the neighbours' degrees. This makes sense when the perspectives are complementary, while the current model assumes each new connection is valuable. However, this case loses strategic complementarity w.r.t.~adding nodes, thus requiring a completely new analysis.

Further possibilities for extending our model include
considering players that minimise the ranking of their degrees among the others,
instead of their own degree. Another option is 
considering a hierarchy of players, where one's possible
strategies and interests depend on her place in the hierarchy.
Another modelling option is to start with a set of connections
on all the graph, rather than only on the non-players, and then
let the players alter their connections while 
attempting not to alter too many connections, since changes can
increase visibility. 

%
In summary, we model sneaky players who aim to 
be connected, and we obtain structural and 
efficiency results that allow us advise the players,
as well as the authority that searches for them.

%% file: omitted_proofs.tex
\section{Omitted Proofs}\label{Sec:omit_proof}

We present the proofs that have been omitted for lack of space.
First, consider the proof of Theorem~\ref{The:ne_not_empt_latt_incl}.
Preparing to prove the theorem, we make an observation and
consider two lemmas about Algorithms~\ref{alg:min_incl_ne}
and \ref{alg:max_incled_ne}.

We first observe that it is enough to consider either additions
or deletions, without adding and deleting simultaneously.
\begin{observation}\label{obs:no_mix_change}
Given edge sets $F \cap G = \emptyset$,
if $u_i((E \cup F) \setminus G) > u_i(E)$, 
at least one of the following holds: 
$u_i(E \cup F) > u_i(E)$ or $u_i(E \setminus G) > u_i(E)$. 
The analogical claim holds for weak improvements.
\end{observation}
\begin{proof}
W.l.o.g., assume both $F$ and $G$ do not intersect $E_0$; if not,
consider $F \setminus E_0$ and $G \setminus E_0$ instead, since
the edges of $E_0$ always exist.

Because of the strategic complementarity and since $F \cap G = \emptyset$,
addition can only decrease the advantage of deletion,
and deletion can only decrease the advantage of addition. Therefore,
$u_i((E \cup F) \setminus G) - u_i(E) \leq
(u_i(E \cup F) - u_i(E)) + (u_i(E\setminus G) - u_i(E))$, and
so $u_i(E \cup F \setminus G) > u_i(E)$ implies
at least one of the following:
$u_i(E \cup F) > u_i(E)$ or $u_i(E \setminus G) > u_i(E)$.
\end{proof}

\begin{lemma}\label{lemma:alg:del_one_enough}
Given graph $G = (V, E)$ and $i \in N$, if $i$ cannot
increase his utility by deleting any single edge,
then he cannot do so by any deletion.
\end{lemma}
\begin{proof}
Assume that $\forall e \in \Neighb(i)$,
$u_i(E \setminus \set{e}) \leq u_i(E)$. Then, by induction,
we prove that $u_i(E \setminus F) \leq u_i(E)$, for any
$F \subseteq \Neighb(i)$, using the strategic complementarity of utility
to edge addition.
\end{proof}

\begin{lemma}\label{lemma:alg:min_incl_ne}
Let $s$ be an inclusion-wise smallest profile resulting in the input graph $G_{in} = (V, E_{in})$.\footnote{In order to obtain such a profile, start from any profiles resulting in $G_{in}$ and delete one by one all the connection actions which deletion does not alter the resulting graph.}
If there are no profitable unilateral deviations from $s$ that delete edges,
then Algorithm~\ref{alg:min_incl_ne} returns the unique
(inclusion-wise) smallest \PANS{} that includes $G_{in}$.

Analogically, if adding edges unilaterally or bilaterally is not profitable in $s$,
then Algorithm~\ref{alg:max_incled_ne} returns the unique
largest \PANS{} included in $G_{in}$.
\end{lemma}
\begin{proof}
We first prove the lemma for Algorithm~\ref{alg:min_incl_ne}.
We assume w.l.o.g.\ that the original strategy profile is a minimal profile
resulting in the given graph, and maintain this throughout the algorithm.
Therefore, unilaterally connecting oneself to another player is never profitable. 

We prove by induction on the loop in
line~\ref{alg:min_incl_ne:imp} that any 
\PANS{} $G' = (V, E')$ that includes $G_{in}$ fulfills $E \subseteq E'$,
so all the added edges are necessary.
The \emph{basis} of the induction, namely the situation before the first
iteration, holds because $G'$ includes the edges of $G_{in}$.
The induction \emph{step} assumes that all the edges in $E$ before
a certain iteration are necessary for an including \PANE, and
considers the edge(s) added next, $F$. Formally, the induction hypothesis means
$E \subseteq E'$ and we need to prove that $E \cup F \subseteq E'$.
Interconnecting non-player neighbours in line~\ref{alg:min_incl_ne:uni_imp:fromS} and maintaining pairwise
stability in line~\ref{alg:min_incl_ne:uni_imp:edge_add} are
clearly necessary for any \PANE{} including E, so it remains
to deal with connecting to non-player neighbours 
in line~\ref{alg:min_incl_ne:uni_imp:fromS:connect_to_nonplayer}.
Assuming by contradiction that $E \cup F \not \subseteq E'$, we conclude that $F \setminus E' \neq \emptyset$.
From the minimality of $T$, adding to $E$ only $F \cap E'$ does not increase
$u_i$. Since $u_i(E \cup F) > u_i(E)$, adding $F \setminus E'$ to $E \cup (F \cap E')$
does increase $u_i$. From the utility comlpementarity to adding edges, this implies that also $u_i(E' \cup (F \setminus E')) > u_i(E')$, 
namely $u_i(E' \cup F) > u_i(E')$, 
contradictory to the stable graph $G'$ having no profitable deviations. Therefore, $E \cup F \subseteq E'$.

Next, we prove by induction on the loop in
line~\ref{alg:min_incl_ne:imp} that no unilateral deletion
is profitable at any step, i.e.~$u_i(E \setminus D) \leq u_i(E), \forall i \in N$,
for any $D \subseteq \Neighb(i)$ at any step of the algorithm.
The induction \emph{basis} follows from the assumption of the lemma that no profitable deletion deviation exists in $G_{in}$,
meaning that~$u_i(E_{in} \setminus D) \leq u_i(E_{in}), \forall i \in N, \forall D \subseteq \set{i} \times N(i)$.
For an \emph{induction step}, we assume by induction that this holds at a certain point
of the execution of Algorithm~\ref{alg:min_incl_ne} and prove
the claim after that point, which adds $F$ to $E$, namely that
$u_i(E \cup F \setminus D) \leq u_i(E \cup F), \forall i \in N, \forall D \subseteq \set{i} \times \Neighb(i)$. Lemma~\ref{lemma:alg:del_one_enough} implies proving this for deleting a single edge will suffice. Now, 
since we know that $u_i(E \setminus D) \leq u_i(E)$, the comlpementarity of adding edges implies that for any $e \in E$,
$u_i(E \cup F \setminus \set{e}) \leq u_i(E \cup F)$ too.
As for the deletion of an edge from $F$, this is clearly
not profitable for the edges added in line~\ref{alg:min_incl_ne:uni_imp:fromS} or in line~\ref{alg:min_incl_ne:uni_imp:edge_add}, and this is also 
not-profitable for edges added in line~\ref{alg:min_incl_ne:uni_imp:fromS:connect_to_nonplayer},
because of the minimality of the added set of edges.

The loop in line~\ref{alg:min_incl_ne:imp} terminates in at most $\abs{V} \choose 2$ steps, since
any step adds edges. By termination,
no addition is profitable, and
as we have just proven, unilateral deletions never become profitable;
thus, Observation~\ref{obs:no_mix_change} implies the graph is Nash stable. It is also \PANS, since
also no bilateral edge addition is possible,
because in particular line~\ref{alg:min_incl_ne:uni_imp:edge_add} is
not applicable now.
The proven above necessity of the added edges implies that
this pairwise Nash stable graph is the minimum possible that includes $G$.

The proof
for Algorithm~\ref{alg:max_incled_ne} is analogous, besides that no
special edge addition like in line~\ref{alg:min_incl_ne:uni_imp:edge_add}
of Algorithm~\ref{alg:min_incl_ne} is required and it is enough
to consider deleting a single edge by player, by
Lemma~\ref{lemma:alg:del_one_enough}. Throughout the algorithm, 
we maintain the assumption that the strategy profile is a minimal profile resulting in the given graph, by deleting an edge from both players' strategies.

More concretely, we first prove by induction that each deletion
is necessary, which is immediate from the complementarity. Next, we
show that at any step, no addition becomes profitable. This can
be inductively demonstrated; in the \emph{basis}, this is assumed.
In the \emph{induction step}, we assume that 
$u_i(E \cup D) \leq u_i(E), \forall D \subseteq N(i)$, and need to conclude that $\forall D \subseteq N(i)$, $u_i(E \setminus\set{e} \cup D) \leq u_i(E\setminus\set{e})$.
If $e \not \in D$, this follows from edge complementarity; otherwise,
it suffices to prove 
$u_i(E \cup D) \leq u_i(E)$ for $D \setminus\set{e}$, which is true
by the induction hypothesis.
Since the deletion process is finite, it terminates, and then, neither addition nor deletion is profitable, so Observation~\ref{obs:no_mix_change} implies the graph is Nash stable.
It is also \PANS, since
also no bilateral edge addition is possible, and this is the largest
\PANS{} included in $G_{in}$.
\end{proof}

We are finally prepared to prove Theorem~\ref{The:ne_not_empt_latt_incl}.

\begin{proof}
We prove the properties of the presented
algorithms one by one.
Algorithm~\ref{alg:least} takes the smallest possible strategy profile, thus allowing for no profitable deviation of deleting edges.
Then, Lemma~\ref{lemma:alg:min_incl_ne} implies 
Algorithm~\ref{alg:min_incl_ne} returns the smallest \PANS{} that can be.
The proof for Algorithm~\ref{alg:great} is analogous, where we start
with the largest possible strategy profile, resulting in the complete
graph.

Next, consider Algorithm~\ref{alg:sup}. It takes the union $G(V, E(s) \cup E(t))$
and applies on it Algorithm~\ref{alg:min_incl_ne}. Since the graphs we unite, namely $G(s)$ and $G(t)$,
are, in particular, Nash stable, strategic complementarity implies there
exist no unilateral deviations of deletion from the union as well.
Therefore, Lemma~\ref{lemma:alg:min_incl_ne} implies Algorithm~\ref{alg:min_incl_ne} returns the unique smallest
\PANS{} that includes the union.
For Algorithm~\ref{alg:inf}, the proof is given by analogy.
\end{proof}

We now prove the complexity Proposition~\ref{prop:run_time}.

\begin{proof}
The number of edges to add is $O((n + m)^2)$, which bounds
the main loop of Algorithm~\ref{alg:min_incl_ne}. Insider that loop,
the loop in line~\ref{alg:min_incl_ne:uni_imp} invokes the
nested loops $O(n \cdot m^2 \cdot 2^m)$ times, and the loop in
line~\ref{alg:min_incl_ne:uni_imp:edge_add} invokes $O(n^2)$ nested loops. 
Since each check and graph manipulation of takes $O((n + m))$,
the total
runtime of Algorithm~\ref{alg:min_incl_ne} is 
$O((n + m)^3 n (n + m^2 \cdot 2^m))$.

As for Algorithm~\ref{alg:max_incled_ne}, its main loop is
bounded by the possible number of edges to delete, namely
$O(n + m)^2$. Inside it, the loop in
line~\ref{alg:max_incled_ne:uni_imp} is in total (in all the iterations of the outer loop)
executed $(On (n + m))$
times, each time taking $O(n + m)$, whereas the second loop
runs $O(m^2)$ times in total, each time using $O(n + n)$. Therefore,
Algorithm~\ref{alg:max_incled_ne} uses $O((n + m)^3)$ time.

Having bounded the core algorithms, we
conclude that the execution time of Algorithms~\ref{alg:least}
and~\ref{alg:sup} is $O((n + m)^3 n (n + m^2 \cdot 2^m))$,
while Algorithms~\ref{alg:great} and \ref{alg:inf} 
run in time $O((n + m)^3 n (n + m^2 \cdot 2^m))$.
\end{proof}

Next, we prove the characterisation Proposition~\ref{prop:ne_char}.

\begin{proof}
Assume $s \in \PANE$. Then, any connection $(i, j) \in E$ of $i\in N$ and $j \not \in N$ implies 
$\deg(j) + \deg(j; i) \geq \alpha_i$, implying part~$1$ of condition~\ref{prop:ne_char:playnoplay}. Additionally, a connection $(i, j) \in E$ between two players $i, j \in N$ implies $\deg(j) \geq \alpha_i, \deg(i) \geq \alpha_j$, namely part~$1$ of condition~\ref{prop:ne_char:playplay}. Moreover, $s \in \PANE$ implies that no unilateral connection building can be profitable, which is part~$2$ of condition~\ref{prop:ne_char:playnoplay}.
Next, pairwise stability implies
condition~\ref{prop:ne_char:playplay}.
As for the last condition, assume $i \in N, j, k \in V \setminus N$,
	$(i, j), (i, k) \in E$. Then, $i$ or another player will connect $j$ and $k$
if $(j, k) \not \in E_0$, since 
$u_i(E \cup \set{j, k}) > u_i(E)$.

Conversely, we now assume that some $s \in S$ satisfies all the conditions above
and demonstrate that $s \in \NE$.
The conditions under number~\ref{prop:ne_char:playnoplay}
imply no player would profit from
a unilateral addition of a set of edges (part~$2$ of number~\ref{prop:ne_char:playnoplay}) or from a deletion (part~$1$ of~\ref{prop:ne_char:playnoplay} and part~$1$ of~\ref{prop:ne_char:playplay}) of edges (there is no profitable deletion of a single edge, thus Lemma~\ref{lemma:alg:del_one_enough} implies no profitable deletion exists altogether).  Condition~\ref{prop:ne_char:noplaynoplay} implies that
no player would gain from changing the connections she makes 
her neighbours have. Then, Observation~\ref{obs:no_mix_change} implies no unilateral change is profitable, namely $\forall i \in N, \forall s'_i \in S_i,
u_i(s'_i, s_{-i}) \leq u_i(s) \Rightarrow s \in NE$.
Condition~\ref{prop:ne_char:playplay} also implies that
$\forall (i, j) \in \set{(k, l) : k, l \in N} \setminus A(s) : 
u_i(s) < u_i(s+(i, j)) \Rightarrow u_j(s) > u_j(s+(i, j))
\Rightarrow s \in \PANE$.
\end{proof}

We need a variation of Observation~\ref{obs:no_mix_change}
to prove Theorem~\ref{The:strong_pane_not_empt_subset_incl} for strong \PANE.
\begin{observation}\label{obs:no_mix_change_set}
Given edge sets $F \cap G = \emptyset$,
if for some $N' \subseteq N$ we have $u_i((E \cup F) \setminus G) \geq u_i(E), \forall i \in N'$, with a strict inequality for at at least one $i \in N'$, then there is $N'' \subseteq N$, such that
one of the following holds for every $i \in N'' \subseteq N$,
with a strict inequality for at at least one $i \in N''$:
$u_i(E \cup F) \geq u_i(E)$ or $u_i(E \setminus G) \geq u_i(E)$. 
\end{observation}
\begin{proof}
W.l.o.g., assume both $F$ and $G$ do not intersect $E_0$; if not,
consider $F \setminus E_0$ and $G \setminus E_0$ instead, since
the edges of $E_0$ always exist.

Because of the strategic complementarity and since $F \cap G = \emptyset$,
addition can only decrease the advantage of deletion,
therefore,
unless there is $N'' \subseteq N$, such that for every $i \in N''$,
$u_i(E \setminus G) \geq u_i(E)$, and $\exists i \in N''$, such that
$u_i(E \setminus G) > u_i(E)$, then
no deletion is profitable after any addition as well. Thus,
$u_i((E \cup F) \setminus G) \geq u_i(E), \forall i \in N'$
and $\exists i \in N'$, where $u_i((E \cup F) \setminus G) > u_i(E)$, implies 
$u_i(E \cup F) \geq u_i(E), \forall i \in N'$
and $\exists i \in N'$, where $u_i(E \cup F) > u_i(E)$.
\end{proof}

We now prove the algorithmic Theorem~\ref{The:strong_pane_not_empt_subset_incl}.

\begin{proof}
Observation~\ref{obs:no_mix_change_set} allows considering additions and deletions separately.
In the $k$-strong case, Algorithm~\ref{alg:min_incl_ne} becomes
Algorithm~\ref{alg:k_min_incl_ne}. There,
we do not take a specific pairwise stability action
of connecting $2$ players, since this is subsumed in a coordinated action
of at least $2$ players in line~\ref{alg:min_incl_ne:uni_imp:fromS:coal_connect:players}.

Having adapted the algorithm in this manner, we first prove by induction
on the loop in line~\ref{alg:k_min_incl_ne:uni_imp} that 
any possible $k$-strong \PANS{} $G' = (V, E')$ including the input
graph $G_{in} = (V, E_{in})$ contains all the added edges, namely $E \subseteq E'$. 
The \emph{basis} of the induction holds, because $E_{in} \subseteq E$.
In the \emph{induction step}, we assume that we have $E \subseteq E'$
at the beginning of a new iteration of line~\ref{alg:k_min_incl_ne:uni_imp}.
The interconnection of the non-player neighbours in 
line~\ref{alg:k_min_incl_ne:uni_imp:fromS:non_pl_neigh} is clearly
necessary on any \PANS{} that includes $E$, so it is left to prove 
that adding some set $F$ in line~\ref{alg:min_incl_ne:uni_imp:fromS:coal_connect} remains in $E'$,
namely $E \cup F \subseteq E'$. 
Assuming by contradiction that $E \cup F \not \subseteq E'$, we conclude that $F \setminus E' \neq \emptyset$.
Now, adding $F \setminus E'$ to $E \cup (F \cap E')$ never decreases
the utility of any player $i \in U \cup W$, since otherwise $i$ would have 
profited from rejecting
that connection, contradicting the statement we will have proven below, namely that
at any step, no deletions are profitable. So we have established that
$\forall i \in U \cup W$,
$u_i((E \cup (F \cap E')) \cup (F \setminus E')) \geq u_i(E \cup (F \cap E'))$,
and then the minimality of $T$ implies that there is some $i \in U \cup W$, for whom the inequality is sharp, namely 
$u_i((E \cup (F \cap E')) \cup (F \setminus E')) > u_i(E \cup (F \cap E'))$.
Now, from the utility comlpementarity to adding edges, this implies that also $u_i(E' \cup (F \setminus E')) \geq u_i(E')$, $\forall i \in U \cup W$,
namely $u_i(E' \cup F) \geq u_i(E')$, 
and $\exists i \in U \cup W$, for whom this inequality is sharp.
This is
contradictory to the stable graph $G'$ having no profitable deviations
of a set of size at most $k$. Therefore, $E \cup F \subseteq E'$.

No deletion was profitable in the original
graph, and we prove by induction that it cannot become profitable during the 
execution of the algorithm,
because we always add a minimum possible set and because of the strategic complementarity.
Formally, the induction \emph{basis} follows from the assumption of the lemma that no profitable deletion deviation exists in $G_{in}$,
meaning that for any subset $D$ of $\cup_{i \in U \cup W}{\set{i} \times N(i)}$,
$u_i(E_{in} \setminus D) \leq u_i(E_{in}), \forall i \in N \Rightarrow u_i(E_{in} \setminus D) = u_i(E_{in}), \forall i \in N $.
For an \emph{induction step}, we assume by induction that this holds at a certain point
of the execution of Algorithm~\ref{alg:k_min_incl_ne} and prove
the claim after that point, which adds $F$ to $E$, namely that
for any subset $D$ of $\cup_{i \in U \cup W}{\set{i} \times N(i)}$,
$u_i(E \cup F \setminus D) \leq u_i(E \cup F), \forall i \in N \Rightarrow u_i(E \cup F \setminus D) = u_i(E \cup F), \forall i \in N$. 
For $D \cap E$, this holds from the induction hypothesis and the complementarity for edge addition.
For $D \cap F$, this holds from the minimality of the added set of edges.
Finally, the complementarity for adding edges implies the induction step for $D$ itself.

Since no deletions become profitable, and no addition in profitable at the end of the algorithm, Observation~\ref{obs:no_mix_change_set}
implies this is a
$k$-strong \PANS. This k-strong \PANS{} is a minimum including one, since all the added edges
are necessary for an including $k$-strong \PANS.
This implies that for each
$k \geq 1$, the $k + 1$-strong \PANS{} constitute a subset of the 
$k$-strong \PANS, and in particular, that a strong pairwise stable graph
exists.

The adaptation of Algorithm~\ref{alg:max_incled_ne} is made
in the loop in line~\ref{alg:max_incled_ne:uni_imp} and proven
analogously, using the ability to unilaterally disconnect.
All the other algorithms stay as they are.
\end{proof}

Next, we prove the general additive bound on efficiency loss,
provided in Proposition~\ref{prop:gen_add_bound}.

\begin{proof}
First, we prove that given any \NE{} profile~$s$, we can find
$t' \in S$, such that $s \subseteq t'$ and
$t' \in \arg\opt_{\hat t \in S}\set{\sw{(\hat t)}}$.
To see this, we take any $t \in \arg\opt_{\hat t \in S}\set{\sw{(\hat t)}}$
and prove that $\sw(s \cup t) \geq \sw(t)$.
Indeed, each $e \in E(s \cup t) \setminus E(t)$ only increases the
social welfare, given the rest of $s \cup t$, since it was profitable
to both its nodes in $s \in NE$, and adding edges keeps that
profitability. Additionally, the edges $E(t)$ yield at least
as much as before adding the edges if $E(s \cup t) \setminus E(t)$.

Now, we shall bound the utility loss because of changing the profile
from any social welfare optimising profile~$t'$
to any Nash equilibrium~$s$ by summing the utility differences from
deleting a sequence of edges between
\begin{inparaenum}[1)]
\item two players,
\item between a player and a non-player
\item and between two non-players.
\end{inparaenum}
%

First, consider the total utility loss $\sw(t') - \sw(s)$ because of
removing an edge $e = (i, j)$ between $2$ players $i, j \in N$.
The loss of $i$'s neighbours other than $j$ is at most
$\deg'_{E(t')}(i) - 1$ (loss of $1$ per a neighbour that loses),
while $i$'s own loss is bounded by $\deg(j) - \alpha_i$. 
Summing up over all the edges in $E(t') \setminus E(s)$ adjacent to $i$,
we obtain (the number neighbours decreases after deleting edges)
$(\deg'_{E(t')}(i) - 1) + (\deg'_{E(t')}(i) - 2) + \dots + 1 = \frac{(\deg'_{E(t')}(i)) (\deg'_{E(t')}(i) - 1)}{2}$ for the
loss of $i$'s neighbours other than $j$, and
$(n - 1)(n + m - 1 - \alpha_i)$
for $i$'s own loss.

Next, consider $i \in N$ disconnecting from some non-players,
possibly rendering those non-players disconnected between themselves.
Here, $i$'s neighbouring players lose up to $(n - 1) m$
(each neighbour loses $m$ because of the decrease in $i$'s degree),
while $i$ should have profited from the disconnection.
Finally, such a disconnection can cost the neighbouring
players of the non-players from which $i$ has disconnected
at most $n \cdot m^2$ (each such player loses $m$ neighbours
of degree at least $m$).

Only players $i$ such that $\alpha_i < n + m - 1$ can have
$u_i(t') > u_i(s)$. Summing over these players yields:
\begin{eqnarray*}
n \frac{(n - 1) (n - 2)}{2}\\ + \abs{\set{i \in N : \alpha_i < n + m - 1}} (n - 1) (n + m - 1)\\
- \set{\sum_{i : \alpha_i < n + m - 1}{\alpha_i}}
+ n ((n - 1) m + n m^2).%
\qedhere
\end{eqnarray*}
\end{proof}

%% file: non_appl_proofs.tex
\section{Non-Applicable Proofs}\label{Sec:non_appl_proofs}

As we mention in \sectnref{Sec:hiders_game:lat_pans}, we prove that the equilibria constitute a lattice with respect
to edge inclusion algorithmically, because 
we cannot conclude this using several known theorems.
To that end, denote \defined{the best response set} of player~$i$ by
$B_i \colon S_{-i} \to 2^{S_i}$, meaning that
$$B_i(s_{-i}) \defas \set{e_i \in S_i \mid 
u_i(e_i, s_{-i}) \geq u_i(e'_i, s_{-i}), \forall e'_i \in S_i}.$$

Next, let \defined{the best-response
correspondence} $B \colon S \to 2^S$ be
$$B(s_1, \ldots, s_n) \defas \set{e \in S \mid
\forall i \in N, e_i \in B_i(s_{-i})}.$$

Firstly, Tarski's theorem~\cite{Tarski1955} is not applicable, since
the best response is, in general, not monotonous, as can be seen in Example~\ref{example:res_graph_pane_dep_strat}. There, $\emptyset 
\subset \set{\cnct(1, 2)}$, but
$B_2(\emptyset) = S_2 \not \subseteq B_2(\set{\cnct(1, 2)}) = \emptyset$.
Secondly, Theorem~$4.1$ by Vives~\cite{Vives1990} claims that if 
\begin{inparaenum}[i)]
\item   \label{enum:vives_cond:B_i_inc}
each $B_i$ increases,
\item   $\forall s \in S$,  $B_i(s_{-i})$ has a smallest element, 
and
\item   $B_i(s_{-i}) \cap \set{e_i \in S_i | e_i \subseteq s_i}$ has a largest element, if it is not empty, 
\end{inparaenum}\label{enum:vives_cond}
then the set of \NE{} is a non-empty complete lattice. In condition~\ref{enum:vives_cond:B_i_inc}, the \defined{increase of $B_i$} means that if $s_j \subseteq s'_j$ for each $j \neq i$, with a strict inclusion for at least one $j \neq i$, then $s_i \in B_i(s_{-i})$ and $s'_i \in B_i(s'_{-i})$ imply that $s_i \subseteq s'_i$. 
Vives' Theorem is also not applicable, as $B_i$ 
is not increasing, as we have shown.
Finally, we could hope to conclude 
that the \NE{} form a non-empty complete lattice from Theorem~$3.1$
in~\cite{JacksonZenou2015} which states that, in a game of
strategic complements, if each player's
best response set is a non-empty closed sublattice of her
strategy set, and its supremum and infimum both non-decrease in the strategies, then the set of Nash equilibria forms a non-empty complete lattice. 
However, in the hiders' game, the suprema and infima of the best responses are not monotonous. Moreover, the game is not a game of strategic complements
w.r.t.~strategies; e.g., in Example~\ref{example:res_graph_pane_dep_strat}, player $2$ would lose from $\cnct(2,1)$ if player~$1$ played $\cnct(1, 2)$, 
while $u_2$ would not change from $\cnct(2,1)$ if $s_1 = \emptyset$.
The game is only a game of \textbf{strategic complements w.r.t.~the actually newly added
edges}, i.e.~$$u_i(E' \cup F) - u_i(E') \geq u_i(E \cup F) - u_i(E),
\forall i \in N, E \subset E', F \cap E' = \emptyset.$$

Furthermore, once we turn to $k$-strong equilibria, those theorems are even more so out of question,
since the $k$-strong best response correspondence
may return the empty set. 
Given the inability to use those theorems, we prove the existence and study the structure of \PANE{} constructively.

%% file: hiders.bbl
\begin{thebibliography}{10}
\expandafter\ifx\csname url\endcsname\relax
  \def\url#1{\texttt{#1}}\fi
\expandafter\ifx\csname urlprefix\endcsname\relax\def\urlprefix{URL }\fi
\expandafter\ifx\csname href\endcsname\relax
  \def\href#1#2{#2} \def\path#1{#1}\fi

\bibitem{ressler2006social}
S.~Ressler, Social network analysis as an approach to combat terrorism: Past,
  present, and future research, Homeland Security Affairs 2~(2) (2006).

\bibitem{chase2002you}
M.~S. Chase, J.~C. Mulvenon, You've got dissent! Chinese dissident use of the
  Internet and Beijing's counter-strategies, Rand Corporation, 2002.

\bibitem{han2018contesting}
R.~Han, Contesting cyberspace in China: Online expression and authoritarian
  resilience, Columbia University Press, 2018.

\bibitem{Choate1991}
P.~Choate, Agents of Influence, Touchstone book, Simon \& Schuster, 1991.

\bibitem{MichalakRahwanWooldridge2017}
T.~Michalak, T.~Rahwan, M.~Wooldridge, Strategic social network analysis
  (2017).

\bibitem{aziz2017weakening}
H.~Aziz, S.~Gaspers, K.~Najeebullah, Weakening covert networks by minimizing
  inverse geodesic length., in: IJCAI, 2017, pp. 779--785.

\bibitem{WaniekMichalakRahwanWooldridge2018}
M.~Waniek, T.~Michalak, T.~Rahwan, M.~Wooldridge, Hiding individuals and
  communities in a social network, Nature Human Behaviour 2 (02 2018).

\bibitem{chen2019ga}
J.~Chen, L.~Chen, Y.~Chen, M.~Zhao, S.~Yu, Q.~Xuan, X.~Yang, Ga-based q-attack
  on community detection, IEEE Transactions on Computational Social Systems
  6~(3) (2019) 491--503.

\bibitem{zhou2019adversarial}
K.~Zhou, T.~P. Michalak, Y.~Vorobeychik, Adversarial robustness of
  similarity-based link prediction, arXiv preprint arXiv:1909.01432 (2019).

\bibitem{zhou2019attacking}
K.~Zhou, T.~P. Michalak, M.~Waniek, T.~Rahwan, Y.~Vorobeychik, Attacking
  similarity-based link prediction in social networks, in: Proceedings of the
  18th International Conference on Autonomous Agents and MultiAgent Systems,
  International Foundation for Autonomous Agents and Multiagent Systems, 2019,
  pp. 305--313.

\bibitem{bottom1984industrial}
N.~R. Bottom, R.~R. Gallati, Industrial espionage: Intelligence techniques and
  countermeasures, Butterworth, 1984.

\bibitem{lindemann2017collaboration}
M.~Lindemann, D.~van Toor, M.~Caianiello, M.~Ferioli, N.~Kovalev, Collaboration
  with justice in the netherlands, germany, italy and canada (2017).

\bibitem{Jackson2003}
M.~O. Jackson, A survey of models of network formation: Stability and
  efficiency, mimeo, in: California Institute of Technology, 2003.

\bibitem{BlochJackson2006}
F.~Bloch, M.~O. Jackson, Definitions of equilibrium in network formation games,
  International Journal of Game Theory 34~(3) (2006) 305--318.

\bibitem{JacksonWolinsky1996}
M.~O. Jackson, A.~Wolinsky, A strategic model of social and economic networks,
  Journal of Economic Theory 71~(1) (1996) 44 -- 74.

\bibitem{GoyalJoshi2006}
S.~Goyal, S.~Joshi, Unequal connections, International Journal of Game Theory
  34~(3) (2006) 319--349.

\bibitem{BalaGoyal2000}
V.~Bala, S.~Goyal, A noncooperative model of network formation, Econometrica
  68~(5) (2000) 1181--1229.

\bibitem{KonigBattistonNapoletanoSchweitzer2011}
M.~D. K{\"o}nig, S.~Battiston, M.~Napoletano, F.~Schweitzer, {Recombinant
  knowledge and the evolution of innovation networks}, Journal of Economic
  Behavior \& Organization 79~(3) (2011) 145 -- 164.

\bibitem{Freeman1978}
L.~C. Freeman, Centrality in social networks conceptual clarification, Social
  Networks 1~(3) (1978) 215 -- 239.

\bibitem{gephi2020}
The open graph viz platform, \url{https://gephi.org/}, accessed on: June, 2020
  (2020).

\bibitem{nodexl2020}
odexl: Network overview, discovery and exploration for excel,
  \url{https://archive.codeplex.com/?p=nodexl}, accessed on: June, 2020 (2020).

\bibitem{ucinet2020}
Ucinet software, \url{https://sites.google.com/site/ucinetsoftware/home},
  accessed on: June, 2020 (2020).

\bibitem{TsvetovatKouznetsov2011}
M.~Tsvetovat, A.~Kouznetsov, Social Network Analysis for Startups: Finding
  connections on the social web, O'Reilly Media, 2011.

\bibitem{TabassumPereiraFernandesGama2018}
S.~Tabassum, F.~S.~F. Pereira, S.~Fernandes, J.~Gama, Social network analysis:
  An overview, WIREs Data Mining and Knowledge Discovery 8~(5) (2018) e1256.

\bibitem{BerlingerioCosciaGiannottiMonrealePedreschi2011}
M.~Berlingerio, M.~Coscia, F.~Giannotti, A.~Monreale, D.~Pedreschi, The pursuit
  of hubbiness: Analysis of hubs in large multidimensional networks, Journal of
  Computational Science 2~(3) (2011) 223 -- 237, social Computational Systems.

\bibitem{KP99}
E.~Koutsoupias, C.~Papadimitriou, Worst-case equilibria, in: 16th Annual
  Symposium on Theoretical Aspects of Computer Science, Trier, Germany, 1999,
  pp. 404--413.

\bibitem{KoutsoupiasPapadimitriou2009}
E.~Koutsoupias, C.~H. Papadimitriou, Worst-case equilibria, Computer Science
  Review 3~(2) (2009) 65--69.

\bibitem{Papadimitriou2001}
C.~Papadimitriou, Algorithms, games, and the internet, in: Proceedings of the
  Thirty-third Annual ACM Symposium on Theory of Computing, STOC '01, ACM, New
  York, NY, USA, 2001, pp. 749--753.

\bibitem{SchulzStierMoses2003}
A.~S. Schulz, N.~Stier-Moses, On the performance of user equilibria in traffic
  networks, in: Proceed.\ of the Fourteenth Annual ACM-SIAM Symp.\ on Discrete
  Algorithms, SODA '03, Society for Industrial and Applied Mathematics,
  Philadelphia, PA, USA, 2003, pp. 86--87.

\bibitem{AnshelevichDasGuptaKleinbergTardosWexlerRoughgarden04}
E.~Anshelevich, A.~DasGupta, J.~Kleinberg, E.~Tardos, T.~Wexler,
  T.~Roughgarden, The price of stability for network design with fair cost
  allocation, in: Foundations of Computer Science, 2004. Proceedings. 45th
  Annual IEEE Symposium on, 2004, pp. 295--304.

\bibitem{Myerson1991}
R.~Myerson, Game Theory: Analysis of Conflict, Harvard University Press, 1991.

\bibitem{DuttavandenNouwelandTijs1998}
B.~Dutta, A.~van~den Nouweland, S.~Tijs, Link formation in cooperative
  situations, International Journal of Game Theory 27~(2) (1998) 245--256.

\bibitem{Myerson1977}
R.~B. Myerson, Graphs and cooperation in games, Mathematics of Operations
  Research 2~(3) (1977) 225--229.

\bibitem{CalvoArmengolIlkilic2009}
A.~Calv{\'o}-Armengol, R.~{\.{I}}lk{\i}l{\i}{\c{c}}, Pairwise-stability and
  nash equilibria in network formation, International Journal of Game Theory
  38~(1) (2009) 51--79.

\bibitem{BelleflammeBloch2004}
P.~Belleflamme, F.~Bloch, Market sharing agreements and collusive networks,
  International Economic Review 45~(2) (2004) 387--411.

\bibitem{JacksonRogers2005}
M.~O. Jackson, B.~W. Rogers, The economics of small worlds, Journal of the
  European Economic Association 3~(2/3) (2005) 617--627.

\bibitem{GoyalJoshi2003}
S.~Goyal, S.~Joshi, Networks of collaboration in oligopoly, Games and Economic
  Behavior 43~(1) (2003) 57 -- 85.

\bibitem{DuttaMutuswami1997}
B.~Dutta, S.~Mutuswami, Stable networks, Journal of Economic Theory 76~(2)
  (1997) 322 -- 344.

\bibitem{JacksonvandenNouweland2005}
M.~O. Jackson, A.~van~den Nouweland, Strongly stable networks, Games and
  Economic Behavior 51~(2) (2005) 420 -- 444, special Issue in Honor of Richard
  D. McKelvey.

\bibitem{Fabrikant2003}
A.~Fabrikant, A.~Luthra, E.~Maneva, C.~H. Papadimitriou, S.~Shenker, On a
  network creation game, in: Proceedings of the Twenty-second Annual Symposium
  on Principles of Distributed Computing, PODC '03, ACM, New York, NY, USA,
  2003, pp. 347--351.

\bibitem{HojmanSzeidl2008}
D.~A. Hojman, A.~Szeidl, Core and periphery in networks, Journal of Economic
  Theory 139~(1) (2008) 295 -- 309.

\bibitem{AumannMyerson1988}
R.~J. Aumann, R.~B. Myerson, Endogenous formation of links between players and
  of coalitions: an application of the Shapley value, Cambridge University
  Press, 1988, pp. 175--192.

\bibitem{GoyalJabbariKearnsKhannaMorgenstern2015}
S.~Goyal, S.~Jabbari, M.~Kearns, S.~Khanna, J.~Morgenstern, Strategic network
  formation with attack and immunization, in: WINE, 2015.

\bibitem{KonigBattistonNapoletanoSchweitzer2012}
M.~D. K{\"o}nig, S.~Battiston, M.~Napoletano, F.~Schweitzer, {The efficiency
  and stability of R\&D networks}, Games and Economic Behavior 75~(2) (2012)
  694 -- 713.

\bibitem{Gallo2012}
E.~Gallo, Small world networks with segregation patterns and brokers, 2012.

\bibitem{Jackson2003b}
M.~O. Jackson, The Stability and Efficiency of Economic and Social Networks,
  Springer Berlin Heidelberg, Berlin, Heidelberg, 2003, pp. 99--140.

\bibitem{BarabasiAlbert1999}
A.-L. Barab{\'a}si, R.~Albert, Emergence of scaling in random networks, Science
  286~(5439) (1999) 509--512.

\bibitem{Tarski1955}
A.~Tarski, A lattice-theoretical fixpoint theorem and its applications.,
  Pacific J. Math. 5~(2) (1955) 285--309.

\bibitem{Vives1990}
X.~Vives, Nash equilibrium with strategic complementarities, Journal of
  Mathematical Economics 19~(3) (1990) 305 -- 321.

\bibitem{JacksonZenou2015}
M.~O. Jackson, Y.~Zenou, Chapter 3 - games on networks, Vol.~4 of Handbook of
  Game Theory with Economic Applications, Elsevier, 2015, pp. 95 -- 163.

\end{thebibliography}
